\documentclass[11pt,oneside]{article}
\usepackage[english]{babel}
\usepackage[utf8]{inputenc}
\usepackage[T1]{fontenc}
\usepackage[letterpaper,top=1in,bottom=1in,left=1in,right=1in,marginparwidth=1in]{geometry}

\usepackage{amsmath,colortbl}
\usepackage{booktabs}
\usepackage{graphicx}
\usepackage[percent]{overpic}
\graphicspath{{./figs/}}
\usepackage{url}
\usepackage[colorinlistoftodos]{todonotes}
\usepackage{bbold}
\usepackage{listings}
\usepackage{xcolor}
\usepackage[framemethod=TikZ]{mdframed}
\usepackage{rotating}
\usepackage{lineno}
\usepackage{tikz}
\usetikzlibrary{arrows.meta, positioning, calc}
\usepackage[most]{tcolorbox}

\newmdenv[
  backgroundcolor=gray!5,
  linecolor=black,
  linewidth=0.4pt,
  leftline=false,
  rightline=false,
  topline=true,
  bottomline=true]{graybox}
\newcommand{\reportnumber}{FERMILAB-PUB-26-0747-CSAID-T}
\usepackage[angle=0,scale=1,color=black,firstpage=true,opacity=1]{background}
\SetBgContents{\footnotesize\sf{\reportnumber}}
\SetBgPosition{current page.north east}
\SetBgHshift{-2in}
\SetBgVshift{-0.5in}
\usepackage{float}
\usepackage[colorlinks=true, allcolors=blue]{hyperref}
\usepackage{placeins}
\usepackage{mathtools}
\usepackage{braket}
\usepackage{slashed}
\usepackage[bottom]{footmisc}
\usepackage{pdfpages}
\usepackage{amsfonts,amsthm,amssymb}
\usepackage{type1cm}
\usepackage{lettrine}
\usepackage{sectsty}
\usepackage{titlesec}
\usepackage{xspace}
\usepackage{multirow}
\usepackage{cleveref}
\usepackage{tabularx}
\numberwithin{equation}{section}
\usepackage{setspace}
\usepackage[parfill]{parskip}
\usepackage{cite}
\definecolor{AgentUserColor}{RGB}{180, 60, 60}
\definecolor{AgentToolColor}{RGB}{200, 100, 0}

\newtcolorbox{agentuser}[1][User]{%
  enhanced, breakable,
  colback=white, colframe=AgentUserColor, boxrule=0.6pt, arc=1.5mm,
  width=\linewidth, top=8pt, bottom=8pt, left=10pt, right=10pt,
  fontupper=\small,
  attach boxed title to top left={yshift=-6pt, xshift=3.5mm},
  boxed title style={colback=white, colframe=white, boxrule=0pt,
    left=0pt, right=0pt, top=3pt, bottom=0pt, sharp corners},
  coltitle=AgentUserColor, fonttitle=\sffamily,
  title={\raisebox{0pt}[0pt][0pt]{#1}},
}

\newtcolorbox{codebox}[1][Code]{%
  enhanced, breakable,
  colback=white, colframe=AgentToolColor, boxrule=0.5pt, arc=1mm,
  width=\linewidth, left=1.5em, right=1em, top=6pt, bottom=6pt,
  fontupper=\small,
  attach boxed title to top left={yshift=-5pt, xshift=3.5mm},
  boxed title style={colback=white, colframe=white, boxrule=0pt,
    left=0pt, right=0pt, top=3pt, bottom=0pt, sharp corners},
  coltitle=AgentToolColor, fonttitle=\sffamily\small,
  title={\raisebox{0pt}[0pt][0pt]{#1}},
}

\newtcolorbox{filecodebox}[1]{%
  enhanced,
  colback=white, colframe=black!45, boxrule=0.4pt, arc=1mm,
  width=\linewidth, left=1.5em, right=1em, top=6pt, bottom=6pt,
  fontupper=\small,
  attach boxed title to top left={yshift=-5pt, xshift=3.5mm},
  boxed title style={colback=white, colframe=white, boxrule=0pt,
    left=0pt, right=0pt, top=3pt, bottom=0pt, sharp corners},
  coltitle=black!65, fonttitle=\ttfamily\small,
  title={\raisebox{0pt}[0pt][0pt]{#1}},
}

\newtcolorbox{codepanel}[1][Code]{%
  enhanced,
  colback=white, colframe=AgentToolColor, boxrule=0.5pt, arc=1mm,
  width=\linewidth, left=1.5em, right=1em, top=6pt, bottom=6pt,
  fontupper=\small,
  attach boxed title to top left={yshift=-5pt, xshift=3.5mm},
  boxed title style={colback=white, colframe=white, boxrule=0pt,
    left=0pt, right=0pt, top=3pt, bottom=0pt, sharp corners},
  coltitle=AgentToolColor, fonttitle=\sffamily\small,
  title={\raisebox{0pt}[0pt][0pt]{#1}},
}

\newtcolorbox{toolcallpanel}[1]{%
  enhanced,
  colback=white, colframe=AgentToolColor, boxrule=0.5pt, arc=1mm,
  width=\linewidth, left=0.8em, right=0.6em, top=5pt, bottom=4pt,
  before skip=5pt, after skip=5pt,
  attach boxed title to top left={yshift=-5pt, xshift=3.5mm},
  boxed title style={colback=white, colframe=white, boxrule=0pt,
    left=0pt, right=0pt, top=3pt, bottom=0pt, sharp corners},
  coltitle=AgentToolColor, fonttitle=\sffamily\small,
  title={\raisebox{0pt}[0pt][0pt]{#1}},
}

\newtcolorbox{auditcodepanel}[1][Code]{%
  enhanced,
  colback=white, colframe=black!45, boxrule=0.5pt, arc=1mm,
  width=\linewidth, left=0.8em, right=0.6em, top=4pt, bottom=4pt,
  fontupper=\small,
  title={},
}

\lstdefinestyle{basecode}{
  basicstyle=\small\ttfamily,
  breaklines=true,
  frame=none,
  numbers=none,
  aboveskip=0pt,
  belowskip=0pt,
  showstringspaces=false,
  keepspaces=true,
  columns=fullflexible,
  upquote=true,
}
\lstdefinestyle{pycode}{style=basecode, language=Python,
  keywordstyle=\color{blue!70!black},
  stringstyle=\color{red!60!black},
  commentstyle=\color{green!45!black}}
\lstdefinestyle{shcode}{style=basecode, language=bash,
  keywordstyle=\ttfamily,
  morekeywords={},
  stringstyle=\color{red!60!black},
  commentstyle=\color{green!45!black}}
\lstdefinestyle{txtcode}{style=basecode, language={}}

\newlength{\trwd}\newlength{\trth}\newlength{\trmid}\newlength{\trclear}
\makeatletter
\newcommand{\trsetdims}{%
  \setlength{\trwd}{\lst@width}%
  \setlength{\trth}{0.45pt}%
  \setlength{\trmid}{0.28\baselineskip}%
  \setlength{\trclear}{0.36\baselineskip}%
}
\makeatother
\newcommand{\trstem}{\rule[-\trclear]{\trth}{\baselineskip}}
\newcommand{\trstemup}{\rule[\dimexpr\trmid-.5\trth\relax]{\trth}%
  {\dimexpr\baselineskip-\trclear-\trmid+.5\trth\relax}}
\newcommand{\trarm}{\rule[\dimexpr\trmid-.5\trth\relax]{2.15\trwd}{\trth}}
\newcommand{\trbox}[2]{{\trsetdims\makebox[#1\trwd][l]{\raisebox{0pt}[0pt][0pt]{%
  \kern\dimexpr.5\trwd-.5\trth\relax #2}}}}
\newcommand{\trvert}{\trbox{1}{\trstem}}
\newcommand{\trtee}{\trbox{3}{\trstem\trarm}}
\newcommand{\trelbow}{\trbox{3}{\trstemup\trarm}}
\lstdefinestyle{treecode}{style=txtcode,
  literate=
    {|}{{\trvert}}{1}
    {|--}{{\trtee}}{3}
    {`--}{{\trelbow}}{3}
}

\newcommand{\nval}{5}

\newcommand{\nllptrials}{40}

\newcommand{\heptabench}{\textsc{HEPTABench}\xspace}
\newcommand{\pkgname}{\textsc{HepBench}\xspace}
\newcommand{\orchestral}{Orchestral\xspace}
\newcommand{\toolbase}{\textsc{ToolBase}\xspace}
\newcommand{\toolbench}{\textsc{ToolBench}\xspace}
\newcommand{\heptapod}{\textsc{HEPTAPOD}\xspace}
\newcommand{\diagrammatica}{\textsc{Diagrammatica}\xspace}

\title{\textbf{\textsc{HEPTABench}: A framework for portable and controlled agentic evaluations in high energy physics}}

\author{Tony Menzo$^{1,2}$\footnote{Corresponding author: \texttt{amenzo@ua.edu}},\, Alexander Roman$^{1}$, George T. Fleming$^{2}$, \\ Sergei Gleyzer$^{1}$, Konstantin Matchev$^{1}$, Stephen Mrenna$^{2}$\\[12pt]
{$^1$}\textit{Department of Physics and Astronomy, University of Alabama,}\\\textit{Tuscaloosa, AL 35487, USA}\\
{$^2$}\textit{Fermi National Accelerator Laboratory, Batavia, IL 60510, USA}
}

\begin{document}
\maketitle
\vspace{-1.1em}

\begin{center}
\begin{minipage}{13cm}
\section*{\centering Abstract}
Quantifying the capability and reliability of agentic systems in high-energy physics (HEP) requires repeated evaluations under controlled configurations. Establishing whether changes to models, tools, or harnesses improve performance also requires these components to be varied independently. We present the \textbf{H}igh \textbf{E}nergy \textbf{P}hysics \textbf{T}ool-and-\textbf{A}gent \textbf{B}enchmarking (\textsc{HEPTABench}) framework for portable and controlled evaluations of agentic systems, with separately configurable tasks, scaffolding, models, providers, harnesses, and domain capabilities. The framework combines HEP benchmark tasks and scientific rubrics with experiment execution and versioned tool interfaces. To support inspection, reproducibility, and post-hoc grading, the framework prioritizes recording and reporting complete experiment specifications, agent trajectories, and experiment artifacts. We demonstrate the framework in two benchmark campaigns. The first campaign evaluates a forward long-lived-particle reach calculation across four frontier coding-agent systems, comparing their built-in coding tools with the same tools supplemented by HEP domain tools. We show that including these HEP domain tools improves numerical accuracy and reduces trial-to-trial dispersion for every tested system. In a complementary campaign covering eight symbolic two-body decay tasks, we show that adding symbolic domain tools increases the fraction of successful trials for Claude Haiku~4.5 from $8\%$ to $79\%$, while GPT-5.5 succeeds in every trial with built-in tools alone. Through systematic evaluation of agentic systems, \heptabench enables the identification of factors that improve the accuracy and reliability of their scientific work.
\end{minipage}
\end{center}

\newpage
\begingroup
\small
\setstretch{1}
\setlength{\parskip}{0pt}
\tableofcontents
\endgroup

\newpage
\section{Introduction}
\label{sec:introduction}

Large language models (LLMs) are now commonly used in executable environments with context management and access to software, data, and computational resources. In these settings, LLMs function as \emph{agents}, planning multi-step tasks, writing and running code, calling tools and invoking external programs, writing and maintaining ``memories'', inspecting intermediate outputs, and revising their methods over successive iterations~\cite{guo2024largelanguagemodelbased,bubeck2023sparks}.
Further supplementing these agents with domain-specific context and tools has enabled demonstrations of agent-assisted science across many fields, including chemistry~\cite{chemcrow,coscientist}, cosmology~\cite{Moss:2025aicosmologist,Xu:2025cmbagent}, exoplanet science~\cite{panek2026asteragenticscience}, quantum computing~\cite{Knipfer:2026kng}, and machine learning for science~\cite{jiang2025agenticscimlcollaborativemultiagentsystems,wei2025aiscienceagenticscience}.
High-energy physics (HEP), in particular, has seen a proliferation of agentic systems spanning collider phenomenology~\cite{Menzo:2025cim,Plehn:2026gxv,ColliderAgent,Agrawal:2026fermiacc,Diefenbacher:2026azr}, experimental analysis, new-physics searches~\cite{Gendreau-Distler:2025fsj,Moreno:2026mqk,Esmail:2026jpb,He:2026jjb,Desai:2026nmx,Diefenbacher:2025zzn},
analysis reproduction and preservation~\cite{Birk:2026zpd,Costa:2026oew}, detector design and experimental operations~\cite{Chung:2026elz,Hill:2026naa,Lugato:2026osi},
symbolic calculations and model building~\cite{Menzo:2026qrl,Gu:2026zbf,Saad:2026pan}, lattice QCD~\cite{Miao:2025sms,Gao:2026cpd}, astroparticle physics, and cosmological phase transitions~\cite{Bakshi:2025fgx,Lucente:2026kgh,Wang:2026jjn}.

These demonstrations now motivate systematic scientific evaluation that quantifies capability and reliability through repeated trials and establishes whether changes to an agentic system improve or impair its performance. Existing community benchmarks provide increasingly realistic tasks for theoretical reasoning, analysis reproduction, and scientific modeling~\cite{Chung:2025nsd,Zhu:2025qnm,Miao:2026gmx,Faroughy:2026dkj,Islam:2026gwb}, as well as input generation for HEP software~\cite{Singh:2026HEPToolBench}, but do not yet provide a common, portable framework for varying individual experimental factors while holding the others fixed~\cite{Menzo:2026persp}. Controlled comparisons require the model, provider, harness, tool and skill loadout, task scaffolding, and execution environment to be separately configurable. Keeping them separate makes it possible to (a) rerun a task through a different harness without rewriting the benchmark, (b) expose the same pinned tool interface to another model, or (c) determine whether a reported difference arose from the model, the tools, or the execution environment. Additionally, existing systems do not consistently support moving an experiment between compatible agent systems or inspecting and reproducing individual trials. Doing so requires an experiment definition that can be executed across compatible agent systems, together with recorded configurations, software environments, agent actions, intermediate outputs, and grading evidence that allow individual trials to be inspected, reproduced, and reassessed.

\begin{figure}[t!]
\centering
\includegraphics[width=0.95\linewidth]{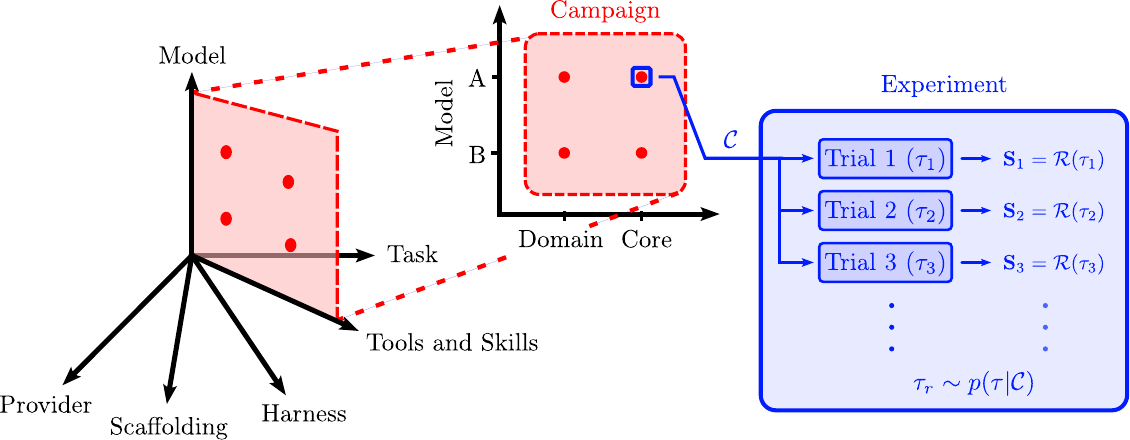}
\caption{Schematic depiction of the six factors defining an agentic evaluation. Left: the shaded slice varies the model and tool-and-skill loadout while holding the other factors fixed. Center: the red points specify four experiments in an illustrative campaign. Right: repeated trials of the highlighted experiment share a fixed configuration $\mathcal C$, sample trajectories $\tau_r$, and receive scientific credits $\mathbf S_r=\mathcal R(\tau_r)$ from the rubric. A campaign comprises the selected configurations and their repeated trials.}
\label{fig:configuration_space}
\end{figure}

{In this paper,} we present \heptabench, a portable framework for controlled evaluation of agentic HEP systems and the software stack that implements it.
The framework treats the task, model, provider, harness, domain tools, and task scaffolding as separately configurable experimental factors, so that the tool interface and the information a task supplies are themselves varied.
Here we refer to an \emph{agentic experiment}\footnote{Not to be confused with a detector-based HEP experiment.} as the evaluation of an agent on a specified physics task under a declared configuration.
Repeated trials of experiments that differ by one factor can associate a change in performance with that factor.
Campaigns with varying factors can measure model--tool, harness--tool, and task--scaffolding interactions.
{ Each experiment
is described once, in a configuration file, and the framework automatically turns that description into individual runs, or \textit{trials} (\cref{fig:configuration_space}).
An experiment can therefore be repeated directly from its recorded specification.
Benchmark tasks and fixed-version tool sets (\textit{pinned loadouts}) can be moved between compatible agent systems without writing custom code for each task.
Because the software environment and the experiment description are both stored alongside the code, anyone can install, check, and rerun the same experiment on a different machine.
Every completed run saves the exact settings used, the agent's step-by-step record of actions, the files it produced, its token usage, and the evidence behind its score.
Results can thus be inspected, and even re-scored under new criteria, without running the model again.}

\heptabench combines three software packages with compatible agent harnesses to form a single declared agentic experiment (see \cref{fig:stack}). Available harnesses and model interfaces include, for example,  \orchestral~\cite{roman2026orchestralaiframeworkagent}, Claude Code~\cite{ClaudeCode}, Codex~\cite{CodexCLI}, OpenCode~\cite{OpenCode}, aisuite~\cite{AISuite}, Cursor~\cite{Cursor}, and Google Antigravity~\cite{GoogleAntigravity}. \toolbase~\cite{toolbase} supplements the harness's built-in tools with versioned domain-tool loadouts, including the HEP-specific operations supplied by \heptapod, and supports their delivery through the Model Context Protocol (MCP)~\cite{MCP}. \toolbench~\cite{toolbench} serves as the experiment \emph{runner}, preparing each trial environment, launching the harness with the selected tool loadout (via \toolbase), grading the submitted outputs, and recording the results. Its dashboard supports viewing campaign results and inspecting individual trials. Built on \toolbench, \pkgname supplies HEP-specific tasks, ground truth, rubrics, and sandbox content in the format required by the runner. Throughout this paper, \heptabench, or ``the \emph{stack},'' refers to the integrated use of \pkgname, \toolbench, \toolbase, and compatible agent harnesses.
External HEP programs must still be installed where required, but their versions and machine-specific paths are resolved outside the benchmark content.

We exercise the framework on two complementary physics benchmarks.
The first study is a reach calculation for the forward production and decay of a long-lived particle (LLP), combining analytic decay widths, forward-meson production, decay-in-flight propagation, detector acceptance, and signal-yield assembly.
For our example muon-philic scalar, the production and decay properties exhibit a threshold behavior, coupling dependence, and channel structure that each provide independent
physical checks to localize errors in a long computational workflow.
The corresponding {benchmark} campaign fixes an out-of-the-box task specification across multiple deployed coding-agent systems and varies the domain loadout.
The HEP domain bundle increases the numerical accuracy and narrows the outcome distribution for every system, while exposing the workflow as a shorter sequence of named, versioned, and human-verified operations that reduces the implementation work required in each trial.
The second study evaluates symbolic decay-rate calculations in generic toy field theories against exact reference answers, with ten repeated trials per task and configuration.
The symbolic campaign compares Claude Haiku~4.5 with and without a symbolic domain-tool loadout on eight shared two-body decay tasks, with GPT-5.5 using only core tools as a baseline.
Grading tests numerical agreement and symbolic equivalence separately.

The remainder of the paper is organized as follows.
\Cref{sec:stack} develops the design principles for reproducible agentic HEP benchmarks and identifies the factors that must be controlled in the experiment.
\Cref{sec:running_example} defines the forward-LLP task and its constituent physics calculations.
\Cref{sec:infrastructure} formulates the calculation as an agent experiment and introduces the software components used to execute it.
\Cref{sec:results} reports the results of the forward-LLP campaign.
\Cref{sec:symbolic_results} reports the results of the complementary symbolic-calculation campaign.
\Cref{sec:discussion} synthesizes the implications and limitations of the two campaigns and concludes.
The LLP reference calculation, performance estimators, and campaign specifications and rubrics are provided in appendices~\ref{app:llp_physics}, \ref{app:metrics}, and~\ref{app:rubrics}.
For readers not interested in the details of the scientific tasks, we suggest reading \Cref{sec:stack}, \Cref{sec:infrastructure} and \Cref{sec:discussion}.

\section{Anatomy of a reproducible agentic evaluation}
\label{sec:stack}

An agentic evaluation grades the scientific outcomes produced when a configured system attempts to complete a specified task.
Interpreting those outcomes requires both a definition of what is measured and a declaration of the factors that determine the agentic experiment.
We first describe the measured scientific outcome and then identify the factors that may be varied or held fixed.

\subsection{Scientific performance as a distribution}

We describe repeated executions at a fixed evaluation configuration $\mathcal C$ by a probability distribution over trajectories.
A physics-aware rubric $\cal R$ for grading an outcome maps each sampled trajectory $\tau$ onto a vector of scientific credits $\mathbf S=(S_1,\ldots,S_M)$,  whose components correspond to graded quantities or stages,
\begin{equation}
    \tau \sim p(\tau\mid\mathcal C),
    \qquad
    \mathbf S = {\cal R}(\tau).
    \label{eq:trajectory_distribution}
\end{equation}
The rubric assigns credit according to the structure of the calculation, accounting for independent components, dependencies on prior steps, and continuous discrepancies such as distances between distributions (see \cref{sec:toolbench_grading}).

Every derived scientific outcome, rubric-stage credit, and overall score is a random variable.
The mean and variance of stage $m$ are
\begin{equation}
    \langle S_m\rangle_{\mathcal C}=\mathbb E[S_m\mid\mathcal C],
    \qquad
    \sigma_m^2(\mathcal C)=\operatorname{Var}(S_m\mid\mathcal C),
    \label{eq:stage_distribution}
\end{equation}
estimated using repeated trials.
The mean measures expected performance on the corresponding scientific component, while the variance measures trial-to-trial consistency.
These quantities must be interpreted together, since low variance may indicate either reliable success or consistent failure, and equal means can conceal different mixtures of outcomes.

The distribution can be interpreted cleanly only if each trial retains the submitted artifacts, grading evidence, trajectory, realized configuration, and operational status.
The record must also pin the versions of external programs, since their results can depend on the version used.
The retained evidence supports checks that distinguish physics failures from infrastructure failures and allows re-grading without new trials.
The corresponding grading and record-keeping mechanisms are described in \cref{sec:toolbench_grading,sec:toolbench_artifacts}.

\subsection{Experimental factors}
\label{sec:factors}

The distribution of scientific outcomes is conditioned on the complete evaluation configuration.
We separate six factors that define this configuration and are commonly entangled in an agentic evaluation (\cref{fig:configuration_space}).
\begin{itemize}
    \item \textbf{Task}: The scientific objective, its fixed input domain, the ground truth, and the rubric.
      The same task can be presented with different amounts of supporting information.

    \item \textbf{Scaffolding}: The task-specific material presented to the agent at the start of a trial: prompt text, supplied formulae and data, input cards, workflow recipes, and the initialized sandbox. Supplying an analytic formula or intermediate distribution changes the scaffolding while leaving the scientific objective and rubric fixed.

    \item \textbf{Model}: The requested LLM, the resolved model snapshot when the provider exposes it, and the inference controls applied to the call, including reasoning mode, sampling settings, and output-token limits.

    \item \textbf{Harness}: The control program that manages the agentic interaction loop. It determines the context policy, control flow, core domain-agnostic operations, termination and intervention behavior, and translation between provider and tool interfaces.
      Harness-level instructions belong to this factor; task-specific instructions belong to scaffolding.

    \item \textbf{Tools and skills}: The domain capabilities made available to the agent. Tools expose executable operations through declared interfaces. Skills may supply instructions, workflow recipes, executable operations, or a combination of these. Instructional content changes the scaffolding; callable operations change the tool interface. These capabilities are distinct from the harness's core operations and are resolved down to the package, component, and version being varied.
    To measure the effect of a tool or skill loadout, we compare conditions with the task, model, provider, and harness held fixed. Any accompanying changes to instructions are part of the loadout comparison and must be recorded as changes to scaffolding. Domain capabilities may also include retrieval tools that search an indexed collection of documents or data and return relevant information as model context, as in retrieval-augmented generation~\cite{Lewis:2020RAG}. The collection, retrieval settings, and returned information should be recorded as part of the experiment.

    \item \textbf{Provider}: The service or local endpoint through which the model is accessed. Providers may differ in model routing, system instructions, sampling implementations, or context handling, even when they advertise the same model. Recording the provider and access mode distinguishes the tested configurations; isolating provider effects requires a campaign designed to compare them.

\end{itemize}

These six factors specify the evaluation configuration for controlled comparisons of agent performance. The execution environment, external-program paths, resource and cost limits, evaluator interventions, and infrastructure status are recorded controls, and enter the experimental design only when deliberately varied. All of these choices define a \emph{condition} in a benchmark campaign. A \emph{trial} is one stochastic execution of a condition, and a \emph{campaign} comprises a set of conditions and repeated trials chosen to measure performance differences or interactions between factors. The following section introduces the forward-LLP task as a working example.

\section{A working example}
\label{sec:running_example}
\begin{figure}[t!]
\centering
\includegraphics[width=1.0\textwidth]{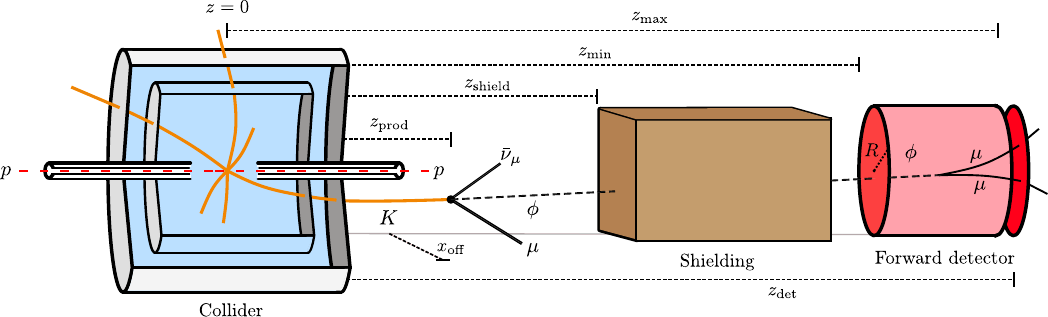}
\caption{Schematic depiction of the forward-LLP geometry and example production and decay channels for the muon-philic scalar model of \cref{sec:llp_target}.
}
\label{fig:forward_llp}
\end{figure}

To demonstrate the framework in a realistic setting, we consider a sensitivity study for a forward long-lived-particle (LLP)~\cite{Alimena:2019zri,Curtin:2018mvb,Ariga:2018uku,Alekhin:2015byh} (fig.~\ref{fig:forward_llp}).
The calculation is representative of many modern phenomenological studies where compact Lagrangians are translated into decay rates and production probabilities, combined with a simulated particle flux, propagated through the fiducial volume of an experiment, and interpreted as observable event yields.
Although each step is standard, the complete study contains many independent calculations that probe common skills associated with phenomenological studies, including symbolic manipulation, Monte Carlo event generation, kinematic sampling, geometric acceptance, and parameter scans with threshold-dependent channels.
This, along with its serial multi-stage structure, makes it well suited for evaluating agent performance.
This analysis is closely related to existing sensitivity studies of the same model at FASER and DarkQuest~\cite{KlingLightScalarsFASER,AbrahamFiegMuonFASER,BlinovMuonScalarDarkQuest}.
Here we isolate production in decays of forward mesons originating from the collider interaction point.
Secondary-muon production in the material of an LHC experimental apparatus and conventional proton-beam-dump production could also have been considered~\cite{AbrahamFiegMuonFASER,RellaMuonphilicProtonBeams,BlinovMuonScalarDarkQuest}; we omit them here to simplify the working example.

\subsection{Muon-philic scalars at forward facilities}
\label{sec:llp_target}
\label{sec:llp_stages}

Consider a real scalar $\phi$ coupled only to muons,\footnote{The interaction in
\cref{eq:llp_lagrangian} should be understood as a low-energy effective coupling after electroweak symmetry breaking. Gauge-invariant completions can generate a muon-enhanced scalar coupling through an extended Higgs sector or through heavy electroweak-charged mediator fields; see, for example, Refs.~\cite{BatellLeptonicHiggsPortal,BatellFlavorSpecificScalars}. The working example does not assume a particular completion.}
\begin{equation}
\mathcal{L}
\;\supset\;
\frac{1}{2}(\partial_\alpha\phi)(\partial^\alpha\phi)
-\frac{1}{2}m_\phi^2\phi^2
\text{$-g\,\phi\,\bar{\mu}\mu$}.
\label{eq:llp_lagrangian}
\end{equation}
Here $\mu$ is the muon field; the scalar mass $m_\phi$ and dimensionless Yukawa coupling $g$ specify the effective model.
The task is to \emph{estimate, for all relevant scalar masses $m_\phi$, the range of couplings, $g$, that produce at least three visible decays inside a downstream forward detector}.
We consider a two-track search for $\phi\to\mu^+\mu^-$ and a collinear-diphoton search for $\phi\to\gamma\gamma$.
The desired result is the expected yield $N_{\rm sig}^{(i)}(m_\phi,g)$ for each search $i\in\{\mu\mu,\gamma\gamma\}$ and the corresponding sensitivity boundary $N_{\rm sig}^{(i)}=3$ in the $(m_\phi,g)$ plane.

A conventional solution proceeds through the following six stages; the formulae and sampling prescriptions are collected in appendix~\ref{app:llp_physics}.

\begin{enumerate}
    \item \textbf{Compute decay rates, branching fractions, and lifetime.}
    Below the dimuon threshold the scalar's leading visible decay is the loop-induced channel $\phi\to\gamma\gamma$, while above threshold the tree-level channel $\phi\to\mu^+\mu^-$ rapidly becomes dominant~\cite{Djouadi:2005gi,BatellFlavorSpecificScalars}.
    Both partial widths are proportional to $g^2$, so one may write $\Gamma_i=g^2\hat\Gamma_i(m_\phi)$.
    The branching fractions are therefore independent of $g$, whereas $c\tau_\phi=\hbar c/[g^2\hat\Gamma_{\rm tot}(m_\phi)]$.

    \item \textbf{Identify the dominant production channels.}
    In the meson-decay channels considered here, $\phi$ is emitted from final-state muons~\cite{AbrahamFiegMuonFASER,BlinovMuonScalarDarkQuest}.
    We include
    \begin{align}
    P&\to\mu\nu_\mu\phi,
    &P&\in\{K,D,D_s\},
    \label{eq:proposed_rewrite_pseudoscalar_channels}\\
    V&\to\mu^+\mu^-\phi,
    &V&\in\{\rho,\omega,\phi(1020),J/\psi,\psi(2S)\}.
    \label{eq:proposed_rewrite_vector_channels}
    \end{align}
    Kaons generally dominate at low mass because of their large forward flux; charm mesons and quarkonia extend the kinematic range as lighter channels close.

    \item \textbf{Generate the forward parent flux.}
    The relevant parent momentum spectra are obtained from inelastic $pp$ collisions at center-of-mass energy $\sqrt{s}$, for example using \textsc{Pythia} with the parents kept stable and harvested before decay~\cite{Bierlich:2022pfr}.
    A forward harvesting angle $\theta_{\max}$ defines the generated flux.
    Event weights normalize the parent flux per inelastic collision using the generated cross section, sample size, and $\sigma_{\rm inel}$. The exposure is fixed by the number of primary interactions $N_{\rm int}$, with $N_{\rm int}=\sigma_{\rm inel}\mathcal L_{\rm int}$ when an integrated luminosity is supplied.

    \item \textbf{Compute and sample the scalar production kinematics.}
    The scalar is attached to each final-state muon line in the Standard Model parent decay amplitude.
    The radiating muon is off shell, so particularly near threshold the rate must be calculated from the full three-body matrix element~\cite{AbrahamFiegMuonFASER,BlinovMuonScalarDarkQuest}.
    For each parent $h$ and scalar mass $m_\phi$, the calculation gives a $g^2$-stripped branching fraction $\hat{\mathcal B}_h(m_\phi)$ and a normalized rest-frame energy spectrum $f_h(x;m_\phi)$, where $x=2E_\phi^\star/m_h$.
    These quantities may either be calculated from the production amplitudes or supplied as tabulated input data.
    Sampling this spectrum and an isotropic direction, followed by a boost with the simulated parent momentum, gives a reusable weighted sample of scalar four-vectors.

    \item \textbf{Propagate through the detector and apply acceptance.}
    Each parent must be propagated until it decays or reaches the absorber at $z_{\rm shield}$, fixing the scalar production vertex $z=z_{\rm prod}$. Parent flight matters for kaons, while charm and vector decays are effectively prompt.
    The scalar trajectory must then be intersected with the cylindrical decay volume specified by $(z_{\min},z_{\max},R,x_{\rm off})$, and its lifetime used to calculate the probability of decay between the entry and exit points.
    The visible decay products must also reach the downstream detector plane at $z_{\rm det}\geq z_{\max}$ within its active radius $R_{\rm det}$. Thus $z_{\min}$ and $z_{\max}$ bound the region where the scalar may decay, while $z_{\rm det}$ specifies where its daughters are detected. All longitudinal coordinates are measured from the interaction point along the beam axis.

    In our benchmark below, we consider purely geometric acceptance, with no detector response, and use the collinear approximation for the diphoton search to test the scalar line of flight against the detector plane.
    The decay point used to evaluate acceptance is fixed at the midpoint of the scalar's path inside the decay volume. This approximation keeps $\mathcal A_j^{(i)}$ independent of $g$, allowing the same sample to be reweighted through the production factor $g^2$ and the lifetime-dependent decay probability.
    The geometry is on axis, $x_{\rm off}=0$, with equal radii for the decay volume and detector plane, $R_{\rm det}=R$.

    \item \textbf{Assemble the yield and determine the sensitivity.}
    Combining the preceding factors gives
    \begin{equation}
    N_{\rm sig}^{(i)}(m_\phi,g)
    =N_{\rm int}\,g^2\mathcal B_i(m_\phi)
    \sum_h\sum_{j\in h} \widehat w_{\phi,j}(m_\phi)
    P_{{\rm dec},j}(m_\phi,g)\mathcal A_j^{(i)},
    \label{eq:proposed_rewrite_signal_yield}
    \end{equation}
    Here $h$ labels the parent species and $j$ runs over the sampled scalars produced from that species; $i$ labels the selected decay channel and is not summed.
    The scalar branching fraction is $\mathcal B_i=\Gamma(\phi\to i)/\Gamma_{\phi,{\rm tot}}$.
    The weight $\widehat w_{\phi,j}$ includes the parent flux normalization and the full production branching fraction with $g^2$ factored out, $\hat{\mathcal B}_h=\Gamma(h\to X\phi)/(g^2\Gamma_{h,{\rm tot}})$, where $X=\mu\nu_\mu$ or $\mu^+\mu^-$; no additional muonic parent branching fraction is required.
    The factors $P_{{\rm dec},j}$ and $\mathcal A_j^{(i)}$ are the probability to decay inside the volume and the search-specific acceptance, respectively.
    At weak coupling $P_{\rm dec}\propto g^2$, and hence $N_{\rm sig}\propto g^4$.
    At large coupling most scalars decay upstream, so the yield decreases exponentially.
    When the peak yield exceeds the event threshold $N_{\rm target}$, the crossings of $N_{\rm sig}^{(i)}(m_\phi,g)=N_{\rm target}$ define the boundaries of the sensitivity region; here $N_{\rm target}=3$.
\end{enumerate}

The Lagrangian and Standard Model particle properties determine the decay physics, but the event yield additionally requires the external inputs introduced above.
The numerical inputs are
\begin{equation}
\mathcal I_{\rm exp}
=
\left\{
\sqrt{s},\ \sigma_{\rm inel},\ N_{\rm int},\ \theta_{\max},\ N_{\rm target}
\right\},
\qquad
\mathcal G
=
\left\{
\text{$z_{\rm shield}$},\ z_{\min},\ z_{\max},\ R,\ z_{\rm det},\ R_{\rm det},\ x_{\rm off}
\right\}.
\label{eq:required_inputs}
\end{equation}
The remaining inputs are a generator/run-card definition of the forward parent flux (or the resulting normalized spectra), the production information $\{\hat{\mathcal B}_h(m_\phi),f_h(x;m_\phi)\}$ for every open channel, and the dimuon and diphoton acceptance definitions.

\begin{figure}[t!]
\centering
\includegraphics[width=0.82\linewidth]{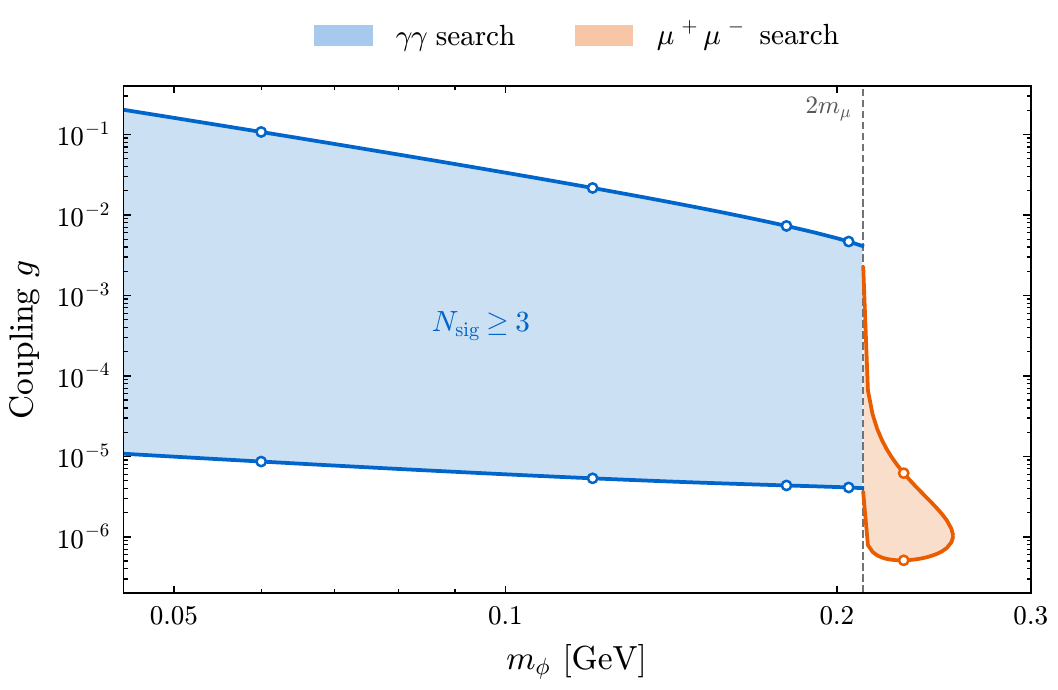}
\caption{Reference sensitivity regions in the scalar mass--coupling plane for the forward-LLP benchmark, using $N_{\rm int}=2.7\times10^{18}$ primary interactions and the geometry and acceptance specified in appendix~\ref{app:llp_physics}. The shaded regions satisfy $N_{\rm sig}\geq3$ under the benchmark's zero-background assumption. The diphoton search is evaluated below the dimuon threshold and the dimuon search above it, as in the benchmark; the dashed line marks $m_\phi=2m_\mu$. The lower and upper boundaries are production-limited and lifetime-limited, respectively. Curves are evaluated on a dense mass grid using the reference calculation, and open circles mark the boundaries at the graded mass points.}
\label{fig:llp_reference_sensitivity}
\end{figure}

The calculation assumes that $\phi\to\mu^+\mu^-$ and $\phi\to\gamma\gamma$ exhaust the scalar width, that the vector parents are unpolarized, and that the supplied production tables accurately represent scalar emission.
An example of the desired output from this calculation is shown in \cref{fig:llp_reference_sensitivity}.
Additional effects relevant to a full experimental analysis, including the deflection of charged particles by magnetic fields~\cite{Ariga:2026MuonSuppression} and detailed particle interactions with detector or shielding material~\cite{Allison:2016lfl,FASER:2024Detector}, are neglected. A realistic experimental analysis would also require trigger and reconstruction efficiencies, detector resolution, backgrounds, and systematic uncertainties.

Above, we specified the target and required inputs for the physics calculation. Defining an agentic experiment also requires specifying what information the agent receives, which calculations it must perform, which tools and skills it can use, and what evidence is retained for grading.

\section{Design and architecture of the benchmarking stack}
\label{sec:infrastructure}

This section describes how the factors introduced in \cref{sec:stack} are represented in the benchmarking stack and carried through an evaluation. The architecture keeps the task specification, supplied scaffolding, domain interface, and deployed agent system separate until an evaluation configuration is resolved. Submitted artifacts (the output files produced by the agent in the sandbox) are then graded against the benchmark rubric, and the realized configuration and trial record are retained for inspection and post-hoc grading.

\begin{figure}[t!]
\centering
\includegraphics[width=0.85\linewidth]{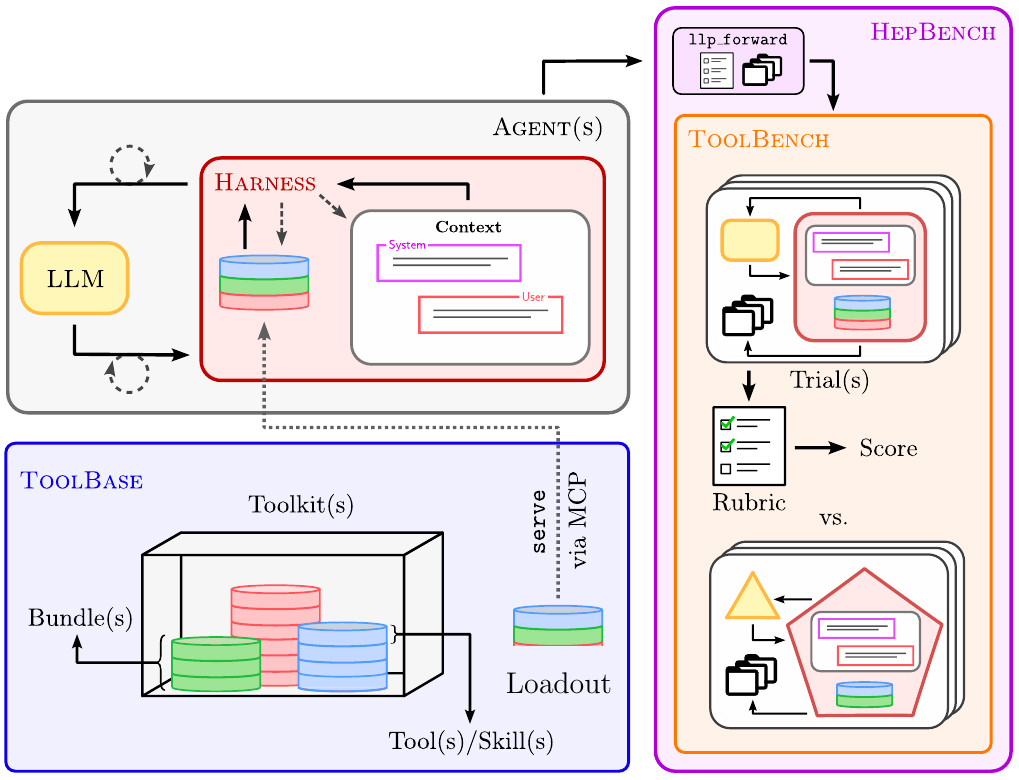}
\caption{Schematic depiction of \emph{the stack}. \pkgname supplies the HEP task and scaffolding; \toolbench resolves and runs the condition; \toolbase serves the declared versioned tool interface; and the harness (Claude Code, Codex, \orchestral, OpenCode) connects that interface to models and providers. Domain toolkits such as \heptapod enter through \toolbase.}
\label{fig:stack}
\end{figure}

\Cref{fig:stack} summarizes the division of labor among the software components. \pkgname stores the task and scaffolding, \toolbench resolves and executes the declared experiment, \toolbase provides the versioned domain capabilities, and the harness runs the agent against the selected model and provider.

\subsection{Declaring the scientific experiment}
\label{sec:toolbench_anatomy}
\label{sec:benchmark_experiment}
\label{sec:benchmark_content}
\label{sec:benchmark_contrasts}

The task and scaffolding factors are encoded in a self-contained \pkgname benchmark directory. A \emph{benchmark} fixes the scientific objective, pinned physical inputs, required output artifacts, ground-truth calculation, and grading rubric.
Only the prompts and inputs available to the agent are copied into the trial sandbox. The ground truth and correctness checks
are unavailable during the trial.

A \emph{variant} fixes the user/system prompts as well as the files placed in the initial sandbox.
Variants retain the same physical objective, ground truth, required output artifacts, and rubric.
Variants differ in the information and guidance supplied to the agent, including analytic expressions, intermediate results, and instructions for using the available software.
These choices determine which parts of the scientific workflow the agent must carry out.
For example, in the forward-LLP benchmark, the scalar widths can be supplied or left for the agent to derive.
Independently, the supplied intermediate inputs define three levels of scaffolding:

\begin{enumerate}
    \item \textbf{Out-of-the-box.}
    The prompt supplies the Lagrangian, the declared width prescription, and the experimental inputs of \cref{eq:required_inputs}. The sandbox is initialized only with the output contract
    (the list of files the agent \textit{must} write and their format), leaving the agent responsible for constructing the complete calculation. We use this variant in the campaigns of \cref{sec:results,sec:symbolic_results}.

    \item \textbf{Production-defined.}
    The sandbox additionally contains the generator cards, stripped production factors $\hat{\mathcal B}_h(m_\phi)$, and normalized spectra $f_h(x;m_\phi)$. The agent generates the parent flux and carries out the downstream propagation and yield calculation.

    \item \textbf{Flux-defined.}
    A validated and normalized forward-parent flux is also supplied. The agent begins with scalar production and remains responsible for the propagation, acceptance, and coupling scan.
\end{enumerate}

Note that these examples shift the starting point within the same scientific workflow without changing the requested result or the grading rubric.
Variants can also modify workflow guidance while retaining the same physics inputs. For example, a variant could add instructions to the system prompt illustrating to the agent how to invoke external software such as \textsc{Pythia}. These instructions help the agent invoke domain-specific software without supplying intermediate physics results or changing the calculations required to complete the task.

An evaluation also specifies a \emph{loadout}, which selects the domain capabilities added to the harness's built-in operations. These capabilities can include validated scientific operations exposed through structured tool interfaces, as well as domain-specific context supplied through skills. The benchmarks below use two recurring classes of loadout. A \emph{core loadout} consists of the harness's domain-independent operations, such as file access, code execution, and task management. A \emph{domain-specific loadout} supplements the core loadout with a declared, versioned set of scientific tools, skills, and associated instructions. Note that each harness maintains its own core interface, so the operations available under a core loadout can differ from one harness to another.

Together, the benchmark, variant, loadout, model, provider, and harness, along with the applicable budgets and model access mode (\textit{e.g.}, metered API, institutional proxy, local endpoint, or subscription-backed application), define an \emph{evaluation configuration}. A campaign selects the configurations to compare and the repeated trials to execute under each. The following subsections describe how the components of the stack translate these declarations into executable trials, graded results, and records that can be inspected and reused.

\subsection{Resolving and executing an evaluation configuration}
\label{sec:toolbench}
\label{sec:toolbench_loop}

In the stack, a scientific experiment specified by an evaluation configuration, as described in the preceding subsection, is realized through a \toolbench-compatible directory containing a top-level specification (in the form of a \texttt{benchmark.yaml} file) and the associated ground-truth reference data, grading checks, variants, harness definitions, and loadouts. Because \toolbench~\cite{toolbench} is a domain-agnostic framework for executing and recording benchmark trials, we maintain the HEP-specific benchmark definitions and scientific inputs in a separate repository, \pkgname. Given a benchmark from \pkgname, \toolbench executes the evaluation configurations selected by a campaign, using the specified trial counts, seeds, and resource limits. It uses \toolbase to resolve the domain loadout and launches the configured harness to run each trial (\cref{sec:toolbase,sec:orchestral}). After execution, \toolbench grades the submitted artifacts, computes performance measures, and retains the trial records for inspection (\cref{sec:toolbench_grading,sec:toolbench_metrics,sec:toolbench_artifacts}). Both campaigns in \cref{sec:results,sec:symbolic_results} use this workflow.

\subsubsection{Versioned domain-tool interfaces}
\label{sec:toolbase}
\label{sec:toolbase_vocab}
\label{sec:orchestral_tools}
\label{sec:toolbase_serve}
\label{sec:toolbase_cli}
\label{sec:toolbase_registry}
Agentic toolkits such as \heptapod provide a large number of tools and skills, of which only a subset may be relevant to a particular task. Selecting these capabilities and configuring the required software can be a laborious process, particularly when the setup must be repeated for different agent harnesses. \toolbase simplifies this process through a common interface for curating and configuring loadouts across compatible harnesses. Domain tools are organized into versioned, publishable collections called \emph{toolkits}. Within a toolkit, related tools can be grouped into \emph{bundles}, while a \emph{loadout} selects the toolkits, bundles, or individual tools to expose through the harness. This allows benchmark authors to curate the available scientific operations without adding toolkit-specific code to the harness or runner. The forward-LLP campaign provides an example of a curated domain-specific loadout (\cref{sec:results_design}).

To supply these tools to the harness, \toolbase identifies the selected implementations and produces the namespaced machine-readable tool schemas exposed to the agent. It uses \orchestral's tool-creation interfaces~\cite{roman2026orchestralaiframeworkagent} to represent the operations as \texttt{BaseTool} instances, which can be supplied directly to a compatible harness or exposed through the Model Context Protocol (MCP)~\cite{MCP}. \orchestral provides the translation from these objects to MCP tool schemas and execution interfaces. Through this common protocol for tool discovery and invocation, the same scientific operations can be made available to different compatible harnesses.

\toolbase also natively supports exporting curated loadouts that specify which tools and toolkit versions to use, allowing the same loadout to be shared across environments and machines. Since external software paths and settings may differ between machines, toolkit authors declare the configuration fields required by their tools. Users can then supply and validate these settings through the \toolbase command-line interface without writing separate configuration scripts. The configuration remains separate from the scientific arguments exposed to the agent, with workspace-dependent paths resolved using the trial sandbox supplied by \toolbench. Together, these features provide \toolbench with a common mechanism for assembling custom loadouts as needed, resolving their declared versions, and serving the selected tools consistently across benchmark trials.

\subsubsection{Harnesses, models, and providers}
\label{sec:orchestral}
\label{sec:orchestral_agents}
\label{sec:orchestral_structured}
To evaluate agentic systems as they are used in scientific workflows, \heptabench supports both general-purpose coding applications and configurable, provider-agnostic harnesses. These include Claude Code~\cite{ClaudeCode}, Codex~\cite{CodexCLI}, OpenCode~\cite{OpenCode}, Cursor~\cite{Cursor}, and Google Antigravity~\cite{GoogleAntigravity}, as well as harnesses built on \orchestral~\cite{roman2026orchestralaiframeworkagent} and aisuite~\cite{AISuite}. The harness manages the interaction loop by supplying prompts and tool schemas to the model, maintaining the conversation state, executing tool calls, applying retry and stopping policies, and returning the final workspace and trajectory. Additional agentic frameworks, including LangChain~\cite{LangChain}, LangGraph~\cite{LangGraph}, Google Agent Development Kit~\cite{GoogleADK}, and Microsoft Agent Framework~\cite{MicrosoftAgentFramework}, provide further options for connecting models to tools and coordinating agent workflows. Here, we focus on supplementing existing agent harnesses with domain-specific tools and context while retaining their native interaction loops. We leave native integration of these additional frameworks into \toolbench to future work.

In addition to the choices that define the evaluation, each harness exposes its own execution settings, such as sandbox permissions, environment variables, and retry policies. These settings must be specified for the selected harness and recorded alongside the other experimental controls. \toolbench accommodates these settings through a harness configuration file that allows it to launch the selected runtime with the declared model access and supported harness-specific options. Where supported by the harness, the configuration can also disable selected built-in tools. For example, the Claude Code configurations in the forward-LLP campaign disable \texttt{WebFetch} and \texttt{WebSearch} to prevent the agent from using these tools to retrieve existing solutions or reference results from the web during a trial.

Models and agent harnesses can be accessed through a number of different methods (for example, via metered APIs, institutional proxies, local endpoints, or subscription-backed applications). Each access mode may expose different inference controls and runtime settings. \toolbench records the selected access mode and uses the harness configuration to specify the applicable controls. For API access, the harness issues the provider request and applies the declared inference settings. Subscription-backed applications allow researchers to evaluate scientific tasks through systems they may already use, but the application may determine the model snapshot, native instructions, context policy, and core tools together. These choices can affect the sequence of actions taken during a trial and must therefore be recorded as part of the evaluation configuration.

\subsection{Grading scientific evidence}
\label{sec:toolbench_grading}

A given scientific task can be graded in many different ways, depending on which outputs are assessed, the criteria used to judge their correctness, and the relative importance assigned to each requirement. In HEP, common scientific outputs include, for example, analytic expressions, intermediate numerical predictions, parameter scans, sensitivity contours, \textit{etc.} In particular, the choice of whether to assess the final result, individual parts of the calculation, or both, and how to credit partially correct work, is largely up to the benchmark author. \toolbench provides explicit conventions for expressing and applying these choices while remaining flexible with respect to the scientific criteria for grading a particular task. A \emph{rubric} records the grading requirements, grouping them into \emph{stages} that can be evaluated separately. Each stage contains one or more \emph{checks}, which inspect the submitted artifacts and test the specified requirements.

Useful grading requirements can concern several aspects of the submitted work, including the following.
\begin{itemize}
    \item \textbf{Presence and usability of outputs:} Producing a requested document, plot, or data file can itself be a requirement of the task, independently of whether its scientific content is correct. \toolbench provides checks for file presence, readability, and required fields, allowing authors to distinguish a missing or unusable submission from an incorrect scientific result.

    \item \textbf{Scientific correctness:} Submitted results may take different forms while satisfying the same physical requirements. Checks can test agreement between algebraically equivalent expressions, expected threshold behavior, or numerical agreement with reference predictions within a stated tolerance. Authors can supplement the built-in checks with custom checking code supplied with the benchmark to evaluate the scientific requirements of their task. These checks evaluate the submitted results independently of the tools used to obtain them, allowing the same rubric to grade an agent that calls \heptapod and one that writes an independent free-form implementation.

    \item \textbf{Degree of numerical agreement:} For distributions, scans, or contours, a pass or fail verdict alone does not capture how closely the submitted result agrees with the reference. A check can therefore also return a \emph{distance} from the reference, labeled with the quantity and units being compared. A benchmark-defined \emph{credit map} converts this distance into continuous \emph{credit} $c\in[0,1]$, the value assigned for scoring. This allows approximate agreement to receive partial credit even when the result falls outside the required tolerance.
\end{itemize}

Each check returns a binary \emph{verdict} indicating whether its requirement was satisfied, together with \emph{grading evidence}, such as the compared values, measured discrepancy, or reason for failure. A stage passes when all of its checks pass. A stage receives either binary credit, one for a pass and zero for a failure, or continuous credit from the specified credit map. Stage \emph{weights} specify the relative contributions to the aggregate score defined in \cref{sec:toolbench_metrics}. Retaining the individual credits allows alternative weights to be examined without repeating the trials.

In some tasks, credit for one result is meaningful only if another requirement has been satisfied. For example, an answer file must be readable before its contents can be assessed, and a benchmark author may require a valid intermediate calculation before awarding credit for a downstream prediction. Other requirements can be evaluated independently, so an incorrect decay width should not automatically remove credit for another correctly computed observable. \toolbench accommodates these choices through \emph{gating} and \emph{non-gating} stages. A gating stage makes its passing verdict a prerequisite for credit from all later stages. A non-gating stage allows later stages to earn credit independently of its verdict. These designations specify how the recorded stage outcomes contribute to the score. They accommodate sequential grading as well as a shared validity gate followed by independently scored quantities, as used in both campaigns reported here.

The reference data, tolerances, credit maps, weights, and gating choices are part of the benchmark's scientific specification. The top-level \texttt{benchmark.yaml} file identifies the benchmark, records the author-specified rubric, and locates its reference data and any custom checking code supplied with the benchmark. The scientific content and grading logic are thus supplied with each benchmark. Comparisons must state explicitly which rubric was applied; the choices for the reported campaigns are given in appendix \ref{app:rubrics}. For these campaigns, \toolbench applies deterministic checks after each trial has ended, keeping reference comparisons out of the agent trajectory. It retains the verdicts, grading evidence, credits, and distances so that the basis of each grade can be inspected. The completed trajectory is also scanned for indications of attempted access to protected reference data. Flagged trials are \emph{quarantined}, receiving zero credit in campaign summaries while retaining the original grade and the evidence that triggered the flag.

\subsection{Trial- and campaign-level performance measures}
\label{sec:toolbench_metrics}

Repeated trials sample the outcome distribution defined in \cref{sec:stack}. Reach combines the stage credits into a single score for each trial, while its distribution across repeated trials characterizes performance under an evaluation configuration. Let $c_{r,m}\in[0,1]$ be the credit earned by trial $r$ at stage $m$, and let $S_{r,m}\in\{0,1\}$ be its threshold verdict.
Denote the number of grading stages by $M$ and the set of gating stages defined in the preceding subsection by $\mathcal{G}$. For nonnegative stage weights $w_m$ with $W=\sum_m w_m>0$, the general per-trial \emph{reach} is
\begin{equation}
  R_r \;=\; \frac{1}{W}\sum_{m=1}^{M} w_m c_{r,m}
  \prod_{\substack{q<m\\ q\in\mathcal{G}}} S_{r,q}\,,
  \label{eq:reach_general}
\end{equation}
where the product is $1$ when no gate precedes stage $m$.
Each stage contributes its weighted credit only if all earlier gates pass. Equivalently, if $m^\ast_r$ is the first gating stage that trial $r$ fails
($m^\ast_r=M$ if none fails), then $\displaystyle R_r = \frac{1}{W}\sum_{m\le m^\ast_r} w_m c_{r,m}$.

For the binary sequential special case, every stage is gating and $c_{r,m}=S_{r,m}$. The definition then reduces to
\begin{equation}
R_r \;=\; \frac{1}{W}\sum_{m=1}^{M} w_m \prod_{q \le m} S_{r,q}\,,
\label{eq:reach}
\end{equation}

so a stage contributes only when it and every preceding stage pass. A task declares success either by requiring all stages to pass or by setting a reach threshold. Across repeated trials, \toolbench reports mean reach together with best-of-$k$ and worst-of-$k$ summaries~\cite{Menzo:2025cim}. Their binary forms are pass@$k$~\cite{chen2021evaluating} and pass$^k$~\cite{yao2024tau}. The estimators, weighting choices, and graded order-statistic generalizations are given in appendix~\ref{app:metrics}.

\subsection{Recording and reproducing the experiment}
\label{sec:toolbench_artifacts}

At the end of the day, the scientific value of a benchmark campaign depends greatly on the ability of an independent researcher to verify its results and reproduce its experimental conditions. This requires clearly identifying those conditions, examining the work performed during each trial, and determining how that work was graded. Campaign records should therefore retain this information for posterity, including the benchmark and rubric revisions, selected evaluation configurations, trial counts, seeds, and resource limits. \toolbench stores these declarations in a \emph{run manifest}, together with the harness and provider settings, model access mode, context policies, and configured evaluator interventions. Since model aliases and loadouts can resolve differently across software revisions or providers, the records also retain the requested configuration and, where available, the resolved model identifiers, toolkit and external-program versions, selected tools, model-visible tool schemas, and available dependencies. This allows researchers to identify differences between the declared configuration and the system that actually executed the trials.

Repeated trials under the same campaign configuration can differ in the actions taken, outputs produced, and reasons for failure. Each trial therefore requires a record of what the agent actually did and produced. \toolbench retains the instructions supplied to the agent, its trajectory, submitted artifacts, grading evidence, and grade, together with the infrastructure and termination status. Operational status is recorded separately from scientific performance so that infrastructure failures, invalid model outputs, early stopping, and scientific failures can be distinguished. To characterize the resources used, trial records also report wall time and, when available, input and output token counts and cache use. Costs are reported for metered APIs, with estimated API-equivalent costs reported for subscription-backed trials.

To audit the scientific work and its evaluation, researchers can use the \toolbench dashboard to move from campaign results to individual trial records, inspect the agent's actions and submitted artifacts, and trace scores to the checks and comparisons that produced them. \toolbench also provides a compact, human-readable summary of each run's results and associated metadata (in a \texttt{summary.txt} file). These records can also be exported for sharing across machines, with trial results and configuration metadata stored in a versioned data format alongside retained answer files, audit logs, and run summaries. Full transcripts can also be included. These exports allow collaborators to inspect the retained evidence and recompute aggregate metrics without access to the original execution environment.

Repeating the experiment on another machine additionally requires the benchmark specification and scientific inputs, compatible software, and the declared model access. The exported loadouts and machine-specific configuration supported by \toolbase (see \cref{sec:toolbase}) allow the selected tools and toolkit versions to be carried over while adapting local software paths and settings. This preserves the specification of the experiment without implying that repeated execution will produce an identical agent trajectory.

Retaining the submitted work also allows authors to revise tolerances, correct a grading check, or reconsider the rubric without repeating model inference. In order for this to be useful, the benchmark's artifact-retention policy must preserve the files required by its grading checks. These artifacts and the recorded trajectories can then be evaluated under a revised deterministic rubric, provided they contain the evidence required by the revised checks. We call this \emph{post-hoc grading} following Ref.~\cite{Menzo:2026persp}, with each revised grade recording the rubric and checking code used.

\section{Forward-LLP benchmark campaign and results}
\label{sec:results}

The first campaign applies the forward-LLP benchmark to four coding-agent systems: Claude Code with Claude Opus~4.8 or Claude Sonnet~5, and Codex with GPT-5.5 or GPT-5.6.
Within each system, the scientific prompt and initial sandbox are held fixed while the supplied domain capabilities are changed.
The campaign measures the accuracy and repeatability of an end-to-end calculation begun without task-specific intermediate results or workflow scaffolding, and records how the implementation presented for audit differs between the two conditions.

\subsection{Experimental design}
\label{sec:results_design}
\label{sec:llp_task}

The scientific task, its six-stage calculation, and the requested sensitivity boundary are defined in \cref{sec:running_example}.
Here both the core and domain arms (to be defined below) use the out-of-the-box variant introduced in \cref{sec:benchmark_experiment}.
Everything needed to define a unique answer is stated in a common user message. The initial sandbox is otherwise empty apart from the output contract, \path{results/answer_schema.json}.

The prompt supplies the Lagrangian and both scalar partial widths, the two production topologies $P\to\mu\nu\phi$ and $V\to\mu^+\mu^-\phi$, the beam energy, generator and tune, the required soft-QCD and charmonium content, forward harvesting cone, inelastic cross section value, exposure, detector geometry, two search acceptances, event threshold, mass grid, probe points, and deliverables.
It therefore fixes the physics target without supplying the calculation that produces it.
In particular, it does not enumerate the contributing parents, provide $\hat{\mathcal B}_h(m_\phi)$ or $f_h(x;m_\phi)$, give a \textsc{Pythia} run card, prescribe how the soft-QCD and
charmonium samples should be combined, specify the displaced parent-decay treatment, provide the ray--cylinder intersection or acceptance algorithms, or choose a Monte Carlo sample size.
Those choices constitute the work being evaluated.

Selected portions of the common user prompt are presented below:

\begin{agentuser}[User]
Estimate the projected signal yield $N_{\rm sig}^{(i)}(m_\phi,g)$ over the $(m_\phi,g)$ plane for a real scalar $\phi$ coupled only to muons, in the forward-detector setup defined below. Treat the dimuon and diphoton searches separately. Use these yields to determine the projected exclusion region for each search, defined by $N_{\rm sig}^{(i)}(m_\phi,g)\geq3$ with zero background. The model is
$$\mathcal L\supset-\tfrac12m_\phi^2\phi^2-g\phi\bar\mu\mu.$$

\begin{center}
\centering
$\vdots$
\end{center}

\textbf{Production:} The scalar is radiated from final-state muons in forward hadron decays, through two topologies,
$$P\to\mu\nu\phi,\qquad V\to\mu^+\mu^-\phi.$$
Which hadrons contribute, and at what rate, is part of the calculation.

\textbf{Beam and flux normalization:} $pp$ collisions at $\sqrt{s}=13.6$ TeV. Generate the forward parent flux with Pythia~8, Monash tune, including soft-QCD light-hadron production and charmonium, and harvest the parent hadrons at production rather than letting the generator decay them. Keep parents within $\theta<0.05$ rad of the beam axis. Take $\sigma_{\rm inel}=80.0$ mb and weight each harvested parent per inelastic collision. The total number of primary interactions is $N_{\rm int}=2.7\times10^{18}$.

\textbf{Apparatus:} All distances are measured from the interaction point along the beam axis, and everything is on-axis.
\begin{itemize}
  \item Parents fly and decay in the open region $0<z<50$ m. Shielding begins at $z=50$ m and absorbs all Standard-Model particles, so a $\phi$ counts only if its parent decayed at $z<50$ m. The $\phi$ itself passes through the shielding freely.
  \item The fiducial decay volume is a cylinder spanning $112<z<163$ m with radius $0.92$ m. The $\phi$ must decay inside it.
  \item The detector is a plane at $z=165$ m with radius $0.92$ m. The visible decay products must reach it.
\end{itemize}

\begin{center}
\centering
$\vdots$
\end{center}
\end{agentuser}

The first omitted passage supplies the exact tree-level dimuon and one-loop diphoton widths, the total width and visible branching fractions, the proper-lifetime normalization, and their $g^2$ scaling.
The final omitted passage defines the dimuon and diphoton searches through purely geometric daughter acceptance at the detector plane, with no magnetic field, material, scattering, efficiency, resolution, or background; it then fixes the 12-point mass grid and the requested width and lifetime probes.

The output contract additionally requires four deliverables: width probes in \path{results/answer.json}, a flat yield scan in \path{results/scan.csv}, the shaded sensitivity region in \path{results/reach.pdf}, and a short report.

Each system is evaluated under two loadouts.
The \textbf{core arm} uses the core loadout defined in \cref{sec:benchmark_experiment}. It retains the shell, file, and \textsc{Python} operations supplied by the installed coding-agent harness and adds no domain-specific capabilities.
The agent must identify and implement the complete production-to-yield pipeline itself.
The \textbf{domain arm} uses the same out-of-the-box variant and augments the harness-native core interface with the pinned \texttt{llp} capability bundle from \heptapod and its accompanying forward-generation skill.
\Needspace{5\baselineskip}
The bundle exposes six typed operations and one accompanying skill:
\begin{codebox}[Bundle (\texttt{llp})]
  \begin{itemize}
    \item \textbf{\textsf{ProductionSpectrum}} computes the reduced parent branching fraction $\hat{\mathcal B}_h(m_\phi)$ and normalized scalar-energy spectrum $f_h(x;m_\phi)$ for a declared parent and scalar mass.
    \item \textbf{\textsf{PythiaFromRunCard}} executes \textsc{Pythia} from a run card and writes the event sample and run manifest.
    \item \textbf{\textsf{HarvestForwardFlux}} selects declared parents within the forward cone and assigns per-inelastic-collision weights.
    \item \textbf{\textsf{MesonDecayToLLP}} combines the parent flux and production spectrum, samples scalar kinematics, and records the displaced parent-decay vertex.
    \item \textbf{\textsf{DecayInVolumeVsCoupling}} propagates the scalar sample through the geometry and returns $N_{\rm sig}(g)$ with lifetime, branching-fraction, and acceptance factors applied.
    \item \textbf{\textsf{DecayInVolumeVsLifetime}} evaluates the same propagation as a function of proper lifetime for validation and reusable lifetime scans.
    \item \textbf{\textsf{PythiaForwardPhysicsRunCardSkill}} supplies versioned guidance on forward \textsc{Pythia} process selection, normalization, and parent harvesting.
  \end{itemize}
\end{codebox}

In preliminary runs, GPT models embedded in Codex did not invoke any of the exposed domain operations without explicit instruction that the domain toolkit was available and should be preferred where applicable.
Because of this, we append a short system-prompt addendum in the Codex domain arm to that effect.
The addendum names no operation, specifies no ordering, and supplies no physics prescription; it is recorded as part of the Codex domain loadout and is absent from the corresponding core arms. 
Claude Code with Opus~4.8 and Sonnet~5 adopted the bundle without additional system-prompt instruction.
Otherwise, the task, user prompt, initial sandbox, model, provider, harness, rubric, seeds, and resource limits are identical in the two arms.
For each of the eight system--loadout pairs we run $n=\nval$ trials, giving \nllptrials{} trials in total. All four systems use subscription-based access, and all eight conditions use the same rubric.

The primary numerical stages use a common log-ratio distance to quantify agreement with the reference.
For a submitted deliverable $X$ (the signal yield $N_{\rm sig}$, lower sensitivity-boundary coupling $g_{\rm lower}$, or upper sensitivity-boundary coupling $g_{\rm upper}$) and reference value $X^{\rm ref}$, one trial has
\begin{equation}
D_X
=
\operatorname{median}_i
\left|
\log_{10}\!\left(\frac{X_i}{X_i^{\rm ref}}\right)
\right|,
\qquad
C_X=10^{-D_X},
\label{eq:llp_distance}
\end{equation}
where $i$ denotes an evaluation point. For example, for $X=N_{\rm sig}$, the 48 evaluation points are the declared $(\text{search},m_\phi,g)$ combinations, with the submitted scan interpolated onto the reference couplings. For $X=g_{\rm lower}$ or $g_{\rm upper}$, each evaluation point is a declared $(\text{search},m_\phi)$ pair for which the reference has a finite sensitivity band; the two boundaries are reconstructed from the submitted scan and evaluated separately.
The distance $D_X$ is reported in decades (dex), where one dex corresponds to a factor of ten, and $C_X$ is the reciprocal of the typical factor by which the calculation differs from the reference. The log-ratio distance treats overestimates and underestimates by the same factor equally, while the median limits the influence of isolated large discrepancies~\cite{Morley:2018LogAccuracy}. This distance and its conversion to credit are choices of the benchmark rubric. Alternative criteria could assess agreement relative to specified uncertainties and correlations, allowing discrepancies to be assigned a statistical interpretation.
The yield and two boundary stages together carry $70\%$ of the reach score, with weights $0.40$, $0.15$, and $0.15$, respectively. The remaining credit comes from widths and lifetime ($0.10$), channel coverage ($0.10$), false reach outside the reference-sensitive region ($0.05$), and the requested plot ($0.05$). The weights are not uniquely determined; here, they prioritize the yield prediction while assigning separate credit to the lower and upper sensitivity boundaries. The same weights are used for all reported conditions; alternative-weight checks are given in appendix~\ref{app:rubrics}. A trial receives no credit unless it supplies valid answer and scan files. Once this structural gate passes, each stage contributes independently; an error in a width probe does not erase credit for an accurate yield scan.
The complete rubric is recorded in appendix \ref{app:rubrics}.
The per-trial reach $R_r$, defined in \cref{eq:reach_general}, combines these stage credits. For a selected pass threshold $R_\star$, a trial is counted as passing when $R_r\geq R_\star$.

\subsection{Results}
\label{sec:results_performance}

\begin{figure}[t!]
\centering
\includegraphics[width=\linewidth]{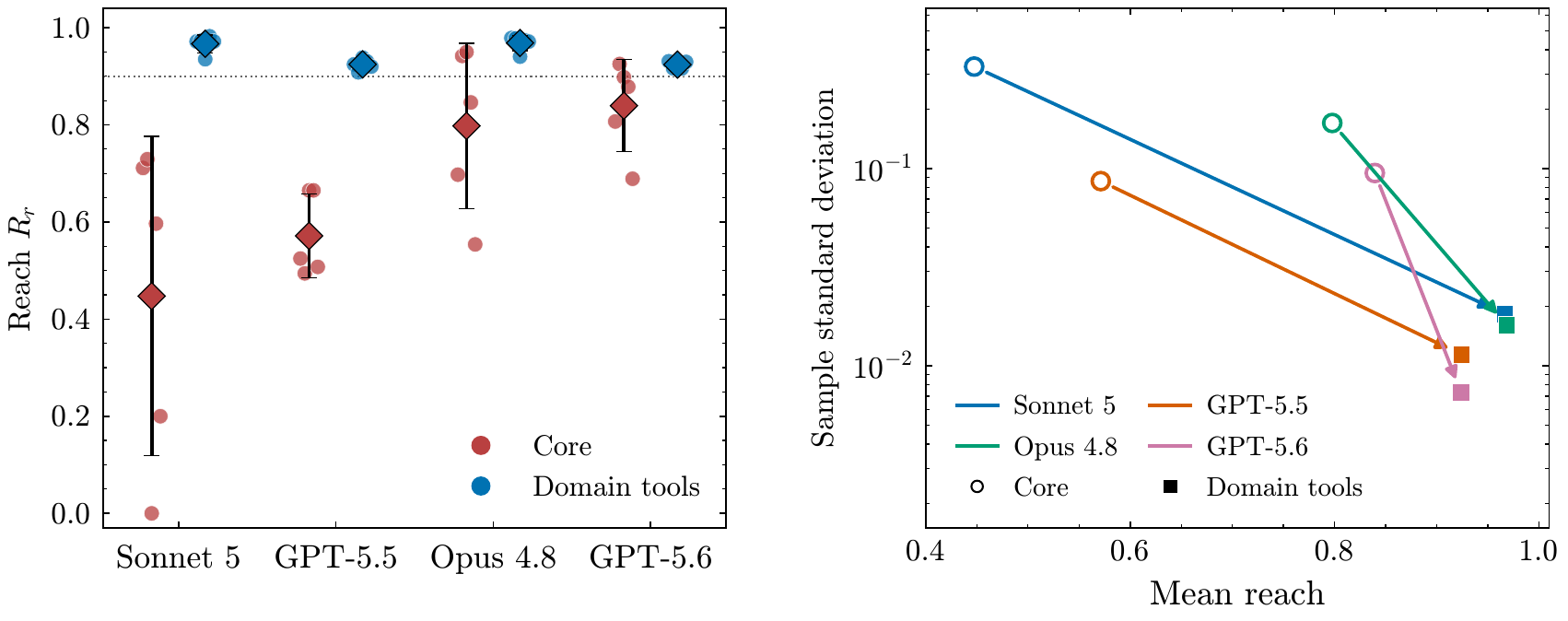}
\caption{Forward-LLP reach for five trials per system and loadout. Left: Individual trials, with diamonds marking sample means and bars showing one sample standard deviation. The dotted line marks $R_\star=0.9$. Right: Mean reach versus sample standard deviation; colors identify systems, and arrows connect core (open circles) to domain (filled squares) conditions. Threshold dependence is shown in \cref{fig:llp_threshold_sensitivity}.}
\label{fig:llp_performance}
\end{figure}

\Cref{fig:llp_performance} summarizes the continuous reach distributions for the eight system--loadout conditions. The left panel shows the five trial scores together with their sample means and standard deviations. For every system, the domain arm has a higher sample mean and a lower sample standard deviation than the core arm. The right panel displays these two changes together. Each arrow moves to the right and downward, corresponding to higher mean reach and less trial-to-trial dispersion in the domain arm.

Reach is continuous, so a binary pass count requires a choice of threshold. The right panel of \cref{fig:llp_threshold_sensitivity} in appendix~\ref{app:campaign_diagnostics} shows how the pass rate changes as the reach threshold $R_\star$ is varied. Because reach combines the stage credits with specified weights, the left panel provides a complementary check using the numerical discrepancies before they are combined into reach. Using the distances defined in \cref{eq:llp_distance}, we define
\begin{equation}
D_{\max}=\max\!\left(D_{N_{\rm sig}},D_{g_{\rm lower}},D_{g_{\rm upper}}\right),
\qquad F_{\max}=10^{D_{\max}},
\label{eq:llp_fmax}
\end{equation}
where missing yields or contour edges are assigned $D_{\max}=\infty$. The maximum is used so that close agreement in one observable cannot compensate for a large discrepancy in another. Consequently, $F_{\max}$ is the largest typical multiplicative discrepancy among the signal yield and the two sensitivity boundaries, and $F_{\max}\leq F$ requires all three to agree with the reference within a factor $F$. At the illustrative value $F=1.5$, this condition is met by $6/20$ core-arm trials and all $20/20$ domain-arm trials. For reach, the campaign takes the representative threshold $R_\star=0.9$, at which $3/20$ core-arm trials and $20/20$ domain-arm trials are counted as passing.

\begin{figure}[t!]
\centering
\includegraphics[width=0.9\linewidth]{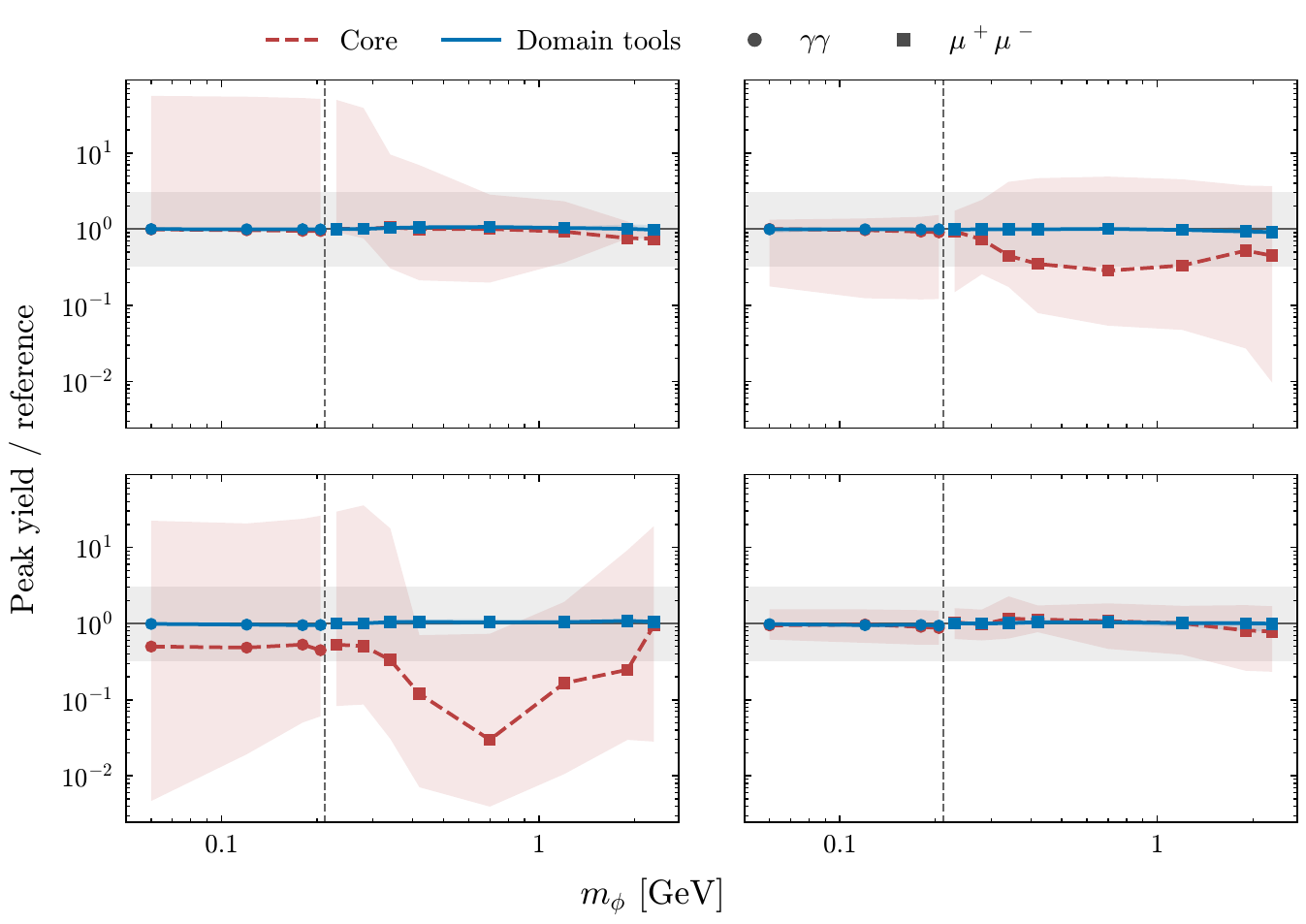}
\caption{Peak signal yield relative to the reference for each forward-LLP system. Panels show Opus~4.8 (upper left), Sonnet~5 (upper right), GPT-5.5 (lower left), and GPT-5.6 (lower right). The curves show the per-mass median peak yield relative to the reference; the shading spans the minimum and maximum values across those trials. Orange dashed curves denote core loadouts and blue solid curves domain loadouts. Circles and squares denote the diphoton and dimuon searches with the vertical dashed line marking the dimuon threshold, $m_\phi=2m_\mu$. All panels share the same scales. The grey band marks agreement within a factor of three. Missing scans and nonpositive peak yields are excluded from these summaries, while all trials remain included in the campaign scores.
}
\label{fig:llp_yields}
\end{figure}

\Cref{fig:llp_yields} compares the submitted peak yields with the reference across the five trials in each system--loadout condition. Comparing the submitted signal yields with the reference across the required masses and couplings gives log-ratio distances of $0.009$--$0.022$ dex in the domain-arm conditions and $0.113$--$0.949$ dex in the core-arm conditions. For comparison, equivalent simulations with 200,000--2,000,000 events give an estimated reach of approximately $0.991$--$0.996$ from sampling fluctuations alone, assuming all other checks pass (see appendix~\ref{app:llp_statistical_reach}). The sensitivity boundaries in \cref{tab:llp_contour} (see appendix~\ref{app:campaign_diagnostics}) show a less uniform response. The domain arm improves the production-limited lower edge for the two Claude Code systems and for GPT-5.5, but not for GPT-5.6, while the lifetime-limited upper edge does not improve consistently.

\Cref{fig:llp_code} shows that the mean amount of agent-authored analysis code is lower by factors of $1.8$--$4.4$ than in the corresponding core-arm conditions. Code volume does not enter the rubric; read together with the observable comparisons and scientific reach, it shows how much of the implementation moves from trial-specific code into named operations in the shared, versioned toolkit. \Cref{fig:llp_audit} compares the numerical procedures recorded in paired Opus~4.8 trials at seed 1002. The domain-arm trajectory represents the calculation as a sequence of named operations, while the core-arm trajectory records the corresponding trial-specific implementation.

Across all four tested systems, the domain configuration gives higher mean reach, less trial-to-trial dispersion, and signal yields closer to the reference calculation. The yield comparison establishes an improvement in numerical accuracy independently of the threshold used to count passing trials (see \cref{fig:llp_threshold_sensitivity} for the threshold dependence), although the sensitivity boundaries do not improve consistently. The agents also write less analysis code, relying on the toolkit for established subcalculations (\cref{fig:llp_code} in appendix~\ref{app:campaign_diagnostics}).

\FloatBarrier
\section{Symbolic benchmark campaign and results}
\label{sec:symbolic_results}

The symbolic campaign compares core and domain-tool loadouts for Claude Haiku~4.5 on eight two-body decay tasks, with GPT-5.5 using core tools as a baseline. Whereas the forward-LLP benchmark combines production, propagation, and acceptance in a single reach calculation, these tasks test the calculation of independent decay widths from specified interaction Lagrangians. Numerical and formula-equivalence checks measure the accuracy of each submitted quantity across ten trials per task and configuration.

\subsection{Experimental design}
\label{sec:symbolic_campaign}

The eight shared tasks cover scalar, Dirac-fermion, and massive-vector decays, using generic fields and deliberately non-round numerical parameters. The core configurations also ran two derivative-coupling tasks that are unsupported by the domain toolset. We exclude those tasks from the comparisons below so that all configurations are evaluated on the same processes.

Each task uses the \texttt{unguided} variant and begins in an empty sandbox. The prompt supplies the interaction Lagrangian, numerical parameters, and averaging and symmetry conventions, together with the units and output format. It provides no reference formula, analogous named process, or worked answer.
The agent must write numerical widths to \texttt{results/answer.json} and, for seven of the eight tasks, supply closed-form expressions using the specified symbols. A line-by-line derivation is requested in \texttt{results/derivation.md} for inspection but does not enter the grade. The excerpt below gives one coupling structure from the scalar-to-fermion task.

\begin{agentuser}[User]
Compute the tree-level partial widths of a real scalar $\phi$ (mass $m_S$) going to fermion pairs, in three separate coupling structures. Couplings are all real.

(a) Scalar Yukawa, $\mathcal{L}=-y\phi\bar\psi\psi$, with $\psi$ a Dirac fermion of mass $m_f$. Compute $\Gamma(\phi\to\psi\bar\psi)$ at $y=0.87$, $m_S=1240$, $m_f=320$ GeV.

\begin{center}
\centering
$\vdots$
\end{center}

Write \texttt{results/answer.json}: widths in GeV under \texttt{part\_a\_width\_gev}, \texttt{part\_b\_width\_gev}, \texttt{part\_c\_width\_gev}, and the closed forms as SymPy-parseable strings under the corresponding \texttt{part\_\{a,b,c\}\_width\_formula} fields. Use symbols \texttt{y,mS,mf} for (a) and (b), and \texttt{yL,yR,mS,m1,m2} for (c).

Record the derivation in \texttt{results/derivation.md}: the conventions adopted, the vertex factors read off the Lagrangian, the amplitude and its spin/polarization sum, the phase-space factor, and the intermediate expressions at each step, in enough detail for the result to be checked line by line. State any limits or consistency checks performed and their outcome.
\end{agentuser}

Haiku~4.5 runs through Claude Code under two loadouts.
The \textbf{core arm} retains the harness-native shell, file, and \textsc{Python} operations and adds no domain-specific capabilities. The agent must derive and evaluate the requested widths with these tools.
The \textbf{domain arm} adds two operations from \heptapod through the \texttt{symbolic-eda} loadout. These operations use the symbolic calculation machinery and the \texttt{SymbolicDiagram} data structure introduced in \diagrammatica~\cite{Menzo:2026qrl}.
\Needspace{5\baselineskip}
\begin{codebox}[Bundle (symbolic tools)]
  \begin{itemize}
    \item \textbf{\textsf{ComputeDecayRate}} computes a tree-level two-body partial width from a declared diagram, returning a closed-form expression and its numerical value for supplied masses and couplings.
    \item \textbf{\textsf{ComputeCrossSection}} computes a tree-level $2\to2$ cross section from declared diagrams, returning a closed-form expression and its numerical value at a supplied center-of-mass energy.
  \end{itemize}
\end{codebox}
Both operations take a diagram specification as input and generate and execute \textsc{FeynCalc} code in a single call. The agent must specify the external particles, their spins and masses, and the vertex structure and couplings, then check that the returned conventions match the requested process. The loadout exposes both operations, although the campaign requests only decay widths.

GPT-5.5 runs through Codex with core tools. All three configurations use subscription access and seeds 1001--1010, giving ten trials per task and 240 trials in total. The Haiku domain arm includes instructions to use the supplied tools for the calculation, and the two Haiku arms use adjacent Claude Code versions. The measured difference therefore includes these changes alongside tool availability. The versions are recorded in appendix~\ref{app:symbolic_details}.

Numerical widths are compared with the reference at order-of-magnitude, $10\%$, and $1\%$ accuracy. Submitted formulae are parsed as \textsc{SymPy} expressions and evaluated at fixed points in the specified parameter domain to test agreement with the reference. This comparison allows algebraically equivalent forms while testing both normalization and parameter dependence. Each requested quantity is checked at the same numerical tolerances, and an error in one quantity does not remove credit for another. A malformed or absent answer file is recorded as a delivery failure. A task inventory, representative rubric, and grading provenance are given in appendix \ref{app:symbolic_details}.

\subsection{Results}
\label{sec:symbolic_performance}

\begin{figure}[t!]
\centering
\includegraphics[width=0.9\linewidth]{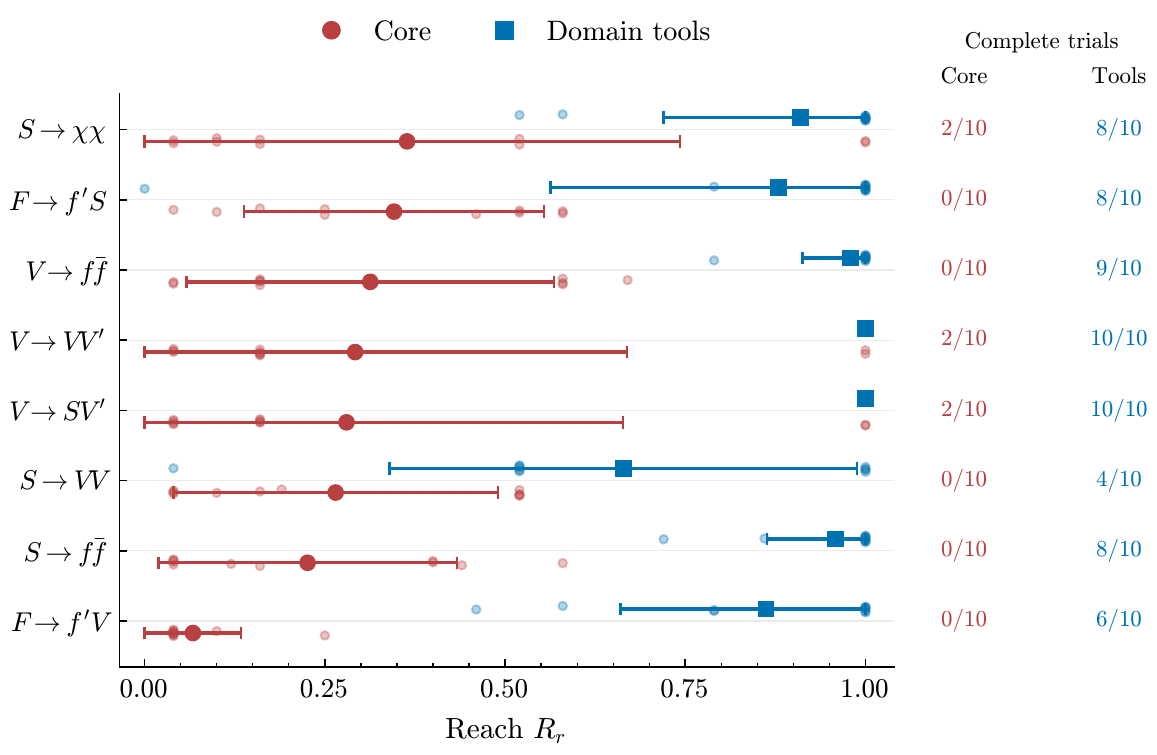}
\caption{Symbolic reach on the eight shared symbolic tasks, ordered by Haiku core mean reach. Faint points show ten individual trials per task and loadout; orange circles and blue squares mark the Haiku core and domain means. Whiskers show one sample standard deviation on either side of the mean, clipped to $0\leq R_r\leq1$. Counts at right give Haiku trials passing every rubric stage.}
\label{fig:symbolic_overview}
\end{figure}

\Cref{fig:symbolic_overview} compares the trial reach distributions for the two Haiku loadouts. The mean reach across the eight tasks increases from $0.2691$ with core tools to $0.9065$ with domain tools. The increase is positive for every task, ranging from $0.40$ to $0.79$. Complete trials increase from $6/80$ to $63/80$. GPT-5.5 completes all $80$ shared-task trials with core tools. With domain tools, Haiku completes every trial for $V\to SV'$ and $V\to VV'$, while $S\to VV$ remains the least reliable task with four complete trials.

\label{sec:symbolic_accuracy}

Haiku core delivers a gradeable answer in all 80 trials, with 74 partial solutions. The domain configuration produces 63 complete solutions, 16 partial solutions, and one trial without a gradeable answer. Among delivered quantities, the fraction satisfying the $1\%$ numerical requirement increases from $32/170$ to $156/168$; formula equivalence increases from $26/150$ to $130/148$. The formula denominator excludes the task for which only numerical answers are requested. GPT-5.5 passes every requested check on the shared tasks.

\label{sec:symbolic_errors}

Among delivered quantities requiring formulae, both checks fail for 122 Haiku core quantities and 12 domain quantities. The numerical width agrees within $1\%$ while the formula fails for two core and six domain quantities. No quantity passes formula equivalence while failing the numerical requirement. Another sixteen core quantities fail the numerical requirement in the task that requests no formula. These checks identify which requirements fail but do not establish the physical cause of an error.

Across the eight shared tasks, the Haiku~4.5 domain configuration improves mean reach, numerical accuracy, and agreement with the reference formulae. The improvement for Haiku suggests that domain tools can broaden the range of models capable of completing these calculations reliably. Although its completion rate remains below the GPT-5.5 baseline, the gain motivates comparing the total cost and execution time of tool-assisted models against more capable models using built-in tools alone. These results are limited to the supported interaction structures tested here, and ten trials per task give limited precision on completion probabilities.

\FloatBarrier
\section{Discussion and conclusions}
\label{sec:discussion}
\label{sec:conclusions}

In this work, we introduced \heptabench, a framework for portable and controlled evaluation of agentic systems in high-energy physics. Its central contribution is the infrastructure to execute benchmarks across compatible agent systems, provide access to versioned toolkits, and inspect and audit individual trials, together with a common specification for defining and sharing physics benchmark tasks and their grading criteria. The framework treats the task, scaffolding, model, provider, harness, and domain capabilities as separately configurable experimental factors, allowing researchers to design comparisons within this six-factor space. By defining the scientific task, supplied inputs, required outputs, and grading criteria independently of the agent implementation, a benchmark author can make the same physics calculation available for evaluation across models, harnesses, and tool interfaces without rewriting the benchmark for each system. Native support for repeated trials allows researchers to quantify accuracy and consistency under each configuration, while retained agent actions, submitted artifacts, and grading evidence allow others to examine how the results were obtained and reassess the conclusions.

To showcase \heptabench, we considered two benchmark campaigns covering a forward-LLP reach calculation and symbolic two-body decay calculations. Within the six-factor space, these campaigns compare standard coding harnesses with and without domain-specific tools and skills across four model--harness systems for the LLP task and one system for the eight symbolic tasks, with a second system providing a core-tool baseline for the latter. For the LLP task, the domain configuration improves numerical accuracy and reduces trial-to-trial dispersion across all four tested systems. For the symbolic tasks, it increases the fraction of Haiku trials passing every rubric stage from $6/80$ to $63/80$, while GPT-5.5 passes every stage with its built-in tools. This improvement suggests that domain tools can broaden the range of viable models for scientific tasks, motivating comparisons of their accuracy, reliability, and computational cost. These comparisons measure complete configurations, including accompanying instructions, and their scope is limited by the trial counts, selected tasks, and shared implementation lineage of the domain tools and reference calculations. The campaigns thus exercise selected comparisons within the six-factor space, while the same specification supports studies of the remaining factors and their interactions.

Looking forward, a natural extension of the platform is to expand the benchmark repository to cover a broader range of theoretical, phenomenological, and experimental workflows, with contributions from the HEP community. Such a repository requires shared conventions for specifying tasks, supplying inputs, defining expected outputs, and grading results, which \heptabench provides while maintaining flexibility for authors to define the physics content and grading criteria appropriate to each task. Researchers can contribute established calculations with validated references and intermediate checks, making their expertise available to others as reusable tests of agent performance. A growing collection of independently authored benchmarks would provide a broader basis for assessing new capabilities, while larger campaigns could establish how reliably improvements transfer across tasks and systems and what computational resources they require. We also leave explicit support for multi-agent benchmarking in \toolbench to future work. By enabling fine-grained control over agentic experiments, providing common conventions for sharing benchmarks, and supporting audits of their results, \heptabench gives a practical basis for determining which advances in agentic systems lead to more accurate and reliable scientific work.

\section*{Public code}
The HEP-specific components of the \heptabench software \textit{stack} are publicly available in the following repositories:
\begin{itemize}
    \item \pkgname contains the benchmark specifications, scientific inputs, grading rubrics, and campaign records.
    \begin{center}
    \url{https://github.com/tonymenzo/hepbench}
    \end{center}
    \item \heptapod contains the scientific tools and skills used in the domain-specific loadouts.
    \begin{center}
    \url{https://github.com/tonymenzo/heptapod}
    \end{center}
\end{itemize}
The stack depends on \toolbench for benchmark execution, grading, reporting, and the dashboard interface, and on \toolbase for selecting, configuring, and serving versioned tool loadouts. The code for these dependencies can be inspected at:
\begin{center}
\begin{tabular}{ll}
\texttt{toolbench}: & \url{https://github.com/tonymenzo/toolbench} \\
\texttt{toolbase}: & \url{https://github.com/alexr314/toolbase}
\end{tabular}
\end{center}

\section*{Acknowledgments}
TM thanks Max Fieg for helpful comments on the forward-LLP calculation.
TM is supported in part by the Shelby Endowment for Distinguished Faculty at the University of Alabama and by Fermilab via Subcontract 725339.
The work of AR and KM is supported in part by the Shelby Endowment for Distinguished Faculty at the University of Alabama and by Fermilab via Subcontract 731293, in support of DOE Award No.\ DE-SCL0000090 ``HEP AmSC IDA Pilot: Knowledge Extraction'' and DOE Award No.\ DE-SCL0000152 ``USQCD AmSC Infrastructure Provision''.
The work of SG and KM is supported in part by the U.S. Department of Energy (DOE) under Award No.\ DE-SC0026347.
This material is based upon work supported by the U.S.~Department of Energy, Office of Science, as part of the Genesis Mission National Science and Technology Challenges under Award Number DE-SC0026840.
This manuscript has been authored by Fermi Forward Discovery Group, LLC under Contract No.~89243024CSC000002 with the U.S. Department of Energy, Office of Science, Office of High Energy Physics.

\newpage
\appendix

\section{Forward-LLP reference calculation}
\label{app:llp_physics}

This appendix supplies the detailed physics and implementation conventions deferred from the main-text presentation in \cref{sec:llp_target}.
It follows the six stages of the reference pipeline: scalar decay, production-channel selection, forward-flux normalization, scalar production and sampling, propagation and acceptance, and final yield assembly.
These equations specify the decay, propagation, and acceptance calculations. Reproducing the reference also requires the recorded production inputs and generator settings.

\subsection{Scalar decay widths and lifetime}
\label{app:llp_decay}

For $m_\phi>2m_\mu$, the tree-level dimuon width is
\begin{equation}
\Gamma_{\mu\mu}(m_\phi,g)
=
\frac{g^2m_\phi}{8\pi}
\left(1-\frac{4m_\mu^2}{m_\phi^2}\right)^{3/2}
\Theta(m_\phi-2m_\mu).
\label{eq:app_width_mumu}
\end{equation}
The $\beta^3$ threshold dependence is characteristic of the scalar current $\bar\mu\mu$; a pseudoscalar current would instead give a $\beta$ dependence.
The same interaction generates a diphoton width through a muon loop~\cite{Djouadi:2005gi,BatellFlavorSpecificScalars},
\begin{equation}
\Gamma_{\gamma\gamma}(m_\phi,g)
=
\frac{\alpha^2g^2m_\phi^3}{256\pi^3m_\mu^2}
\left|A_{1/2}(\tau)\right|^2,
\qquad
A_{1/2}(\tau)
=
\frac{2\left[\tau+(\tau-1)f(\tau)\right]}{\tau^2},
\label{eq:app_width_gammagamma}
\end{equation}
where $\tau=m_\phi^2/(4m_\mu^2)$ and
\begin{equation}
f(\tau)=
\begin{cases}
\arcsin^2\!\sqrt{\tau}, & \tau\leq1,\\[3pt]
-\dfrac14
\left[
\ln\!\left(
\dfrac{1+\sqrt{1-\tau^{-1}}}
      {1-\sqrt{1-\tau^{-1}}}
\right)-i\pi
\right]^2, & \tau>1.
\end{cases}
\label{eq:app_loop_function}
\end{equation}
In particular, $A_{1/2}\to4/3$ for $\tau\to0$.
The tree-level expression is checked against an explicit Dirac-spinor calculation, and the loop width against an independent Feynman-parameter evaluation.

In the minimal model these are the only decay channels, so
\begin{equation}
\Gamma_{\phi,{\rm tot}}
=\Gamma_{\mu\mu}+\Gamma_{\gamma\gamma},
\qquad
\mathcal B_i=\frac{\Gamma_i}{\Gamma_{\phi,{\rm tot}}},
\qquad
c\tau_\phi=\frac{\hbar c}{\Gamma_{\phi,{\rm tot}}}.
\label{eq:app_lifetime}
\end{equation}
Both amplitudes contain one power of $g$.
It is therefore useful to write $\Gamma_i=g^2\hat\Gamma_i(m_\phi)$: the branching fractions are independent of $g$, while $c\tau_\phi=\hbar c/[g^2\hat\Gamma_{\rm tot}(m_\phi)]$.

\subsection{Production channels and three-body rates}
\label{app:llp_production}

The scalar is emitted from final-state muons in two families of forward-hadron decays,
\begin{align}
P&\to\mu\nu_\mu\phi,
&P&\in\{K,D,D_s\},
\label{eq:app_pseudoscalar_channels}\\
V&\to\mu^+\mu^-\phi,
&V&\in\{\rho,\omega,\phi(1020),J/\psi,\psi(2S)\}.
\label{eq:app_vector_channels}
\end{align}
The pion channel is kinematically closed over the mass range studied.
For pseudoscalar parents the underlying weak amplitude is fixed by $G_F$, the relevant CKM element, and the meson decay constant.
For vector parents the effective dimuon coupling is normalized to the measured $V\to\mu^+\mu^-$ partial width; the two diagrams in which $\phi$ is attached to either muon leg interfere.
Explicit amplitudes may be found in Refs.~\cite{CarlsonRislowMesonDecay,MitraSahooVectorDecay}, and their application to forward scalar production is summarized in Ref.~\cite{AbrahamFiegMuonFASER}.

The radiating muon is internal and off shell, with momentum $q=p_\mu+p_\phi$ and propagator
\begin{equation}
S_F(q)=\frac{i(\slashed q+m_\mu)}{q^2-m_\mu^2},
\qquad
q^2-m_\mu^2=m_\phi^2+2p_\mu\!\cdot p_\phi.
\label{eq:app_offshell_muon}
\end{equation}
Writing $X=\mu\nu_\mu$ for a pseudoscalar parent and $X=\mu^+\mu^-$ for a vector parent, the differential rate is
\begin{equation}
d\Gamma_h
=
\frac{1}{2m_h}\overline{|\mathcal M_h|^2}\,d\Phi_3.
\label{eq:app_production_rate}
\end{equation}
The full three-body expression is required near kinematic thresholds; multiplying an ordinary muonic branching fraction by an on-shell splitting probability does not reproduce the off-shell rate.
As validation, the vector amplitude satisfies the appropriate Ward identity to numerical precision and reproduces the measured two-body normalization when the scalar-emission vertex is removed.

Integrating \cref{eq:app_production_rate} gives the partial width and hence
\begin{equation}
\mathcal B_h(g,m_\phi)
\equiv
\frac{\Gamma_h(g,m_\phi)}{\Gamma_{h,{\rm tot}}}
=g^2\hat{\mathcal B}_h(m_\phi).
\label{eq:app_production_branching}
\end{equation}
The same matrix element defines the normalized scalar-energy distribution in the parent rest frame,
\begin{equation}
f_h(x;m_\phi)
=
\frac{1}{\Gamma_h}\frac{d\Gamma_h}{dx},
\qquad
x=\frac{2E_\phi^\star}{m_h},
\qquad
\int_{\rm kin.}dx\,f_h(x;m_\phi)=1.
\label{eq:app_x_spectrum}
\end{equation}
The recoil leptons are integrated out.
Both $\Gamma_h$ and $d\Gamma_h/dx$ are proportional to $g^2$, so $\hat{\mathcal B}_h$ and $f_h$ need be computed only once per scalar mass.

For a spin-zero parent, rotational invariance makes the scalar direction isotropic after the recoil system has been integrated over.
The same prescription applies to an unpolarized ensemble of vector parents, which is an explicit modeling assumption here:
\begin{equation}
\frac{d\Gamma_h}{dE_\phi^\star d\Omega^\star}
=
\frac{d\Gamma_h}{dE_\phi^\star}\frac{1}{4\pi}.
\label{eq:app_isotropy}
\end{equation}
After drawing $x$ from \cref{eq:app_x_spectrum} and an isotropic direction, the scalar four-vector is boosted to the laboratory frame using the simulated parent momentum. Only this scalar four-vector is retained for the inclusive sensitivity calculation.

\subsection{Forward parent flux and event weights}
\label{app:llp_flux}

The parent spectra are generated in inelastic $pp$ collisions at $\sqrt{s}=13.6$~TeV.
One implementation uses an unweighted soft-QCD sample from \textsc{Pythia}~\cite{Bierlich:2022pfr}, keeps the relevant parents stable, and records each parent momentum before decay.
If the sample contains $N_{\rm gen}$ inelastic events, each harvested occurrence $j$ of species $h$ carries weight
\begin{equation}
w_{h,j}=\frac{1}{N_{\rm gen}},
\qquad
\sum_{j\in h}w_{h,j}=\langle n_h\rangle,
\label{eq:app_parent_weight}
\end{equation}
so the sum gives the mean multiplicity of $h$ per inelastic collision. For a separately generated process sample $a$ with cross section $\sigma_a$ and $N_a$ generated events, the corresponding weight is $\sigma_a/(N_a\sigma_{\rm inel})$, before any sampling corrections. Combining process samples requires a prescription that avoids double counting overlapping production. The reference production inputs retain these weights; the weight $1/N_{\rm gen}$ in \cref{eq:app_parent_weight} applies only to an inclusive sample normalized to $\sigma_{\rm inel}$.
For integrated luminosity $\mathcal L_{\rm int}$ and pinned inelastic cross section $\sigma_{\rm inel}$, the number of primary collisions is
\begin{equation}
N_{\rm int}=\sigma_{\rm inel}\mathcal L_{\rm int}.
\label{eq:app_collider_normalization}
\end{equation}

If $n_h$ daughter throws are made for a harvested parent, the coupling-independent weight assigned to each scalar is
\begin{equation}
\widehat w_{\phi,j}
=
w_{h,j}\frac{\hat{\mathcal B}_h(m_\phi)}{n_h}.
\label{eq:app_scalar_weight}
\end{equation}
Its physical production weight at coupling $g$ is $g^2\widehat w_{\phi,j}$.
The sum includes every kinematically open parent and, through $\hat{\mathcal B}_h$, both interfering muon-leg diagrams for vector parents.

\subsection{Parent propagation and detector geometry}
\label{app:llp_geometry}

The scalar is produced at the parent decay vertex.
For a parent with momentum $\mathbf p_{h,j}$ and proper decay length $c\tau_h$, the mean laboratory-frame decay length is
\begin{equation}
\lambda_{h,j}
=\beta_{h,j}\gamma_{h,j}c\tau_h
=\frac{|\mathbf p_{h,j}|}{m_h}c\tau_h.
\label{eq:app_parent_decay_length}
\end{equation}
The shielding is represented by an absorber beginning at $z=z_{\rm shield}$, the last allowed longitudinal position of a parent decay.
For a forward parent produced at $\mathbf r_{h,j}^{\rm prod}$, its maximum flight distance before reaching the absorber is
\begin{equation}
\ell_{h,j}^{\rm abs}
=
\frac{\text{$z_{\rm shield}$}-z_{h,j}^{\rm prod}}
     {\widehat p_{h,j,z}}.
\label{eq:app_parent_absorber_distance}
\end{equation}
One may draw $\ell_{h,j}$ from an exponential distribution of mean $\lambda_{h,j}$ and retain the decay only for $0\leq\ell_{h,j}<\ell_{h,j}^{\rm abs}$.
The scalar production point is then
\begin{equation}
\mathbf r_{0,j}
=\mathbf r_{h,j}^{\rm prod}
+\ell_{h,j}\widehat{\mathbf p}_{h,j}.
\label{eq:app_parent_decay_vertex}
\end{equation}
Equivalently, one may sample from the exponential distribution truncated at $\ell_{h,j}^{\rm abs}$ and multiply the event weight by $1-\exp(-\ell_{h,j}^{\rm abs}/\lambda_{h,j})$.
This treatment matters for kaons, which travel macroscopic distances; charm mesons and vector mesons are effectively prompt.

The detector is modeled as an on-axis cylindrical decay volume
\begin{equation}
\mathcal V
=
\left\{(x,y,z):
z_{\min}\leq z\leq z_{\max},\quad
(x-x_{\rm off})^2+y^2\leq R^2
\right\},
\label{eq:app_decay_volume}
\end{equation}
followed by a circular detector plane of radius $R_{\rm det}$ at $z=z_{\rm det}\geq z_{\max}$. The interval $[z_{\min},z_{\max}]$ defines where the scalar may decay, and the plane at $z_{\rm det}$ defines the surface that its visible decay products must reach.
The geometry is thus specified by $(z_{\rm shield},z_{\min},z_{\max},R,z_{\rm det},R_{\rm det},x_{\rm off})$; the on-axis configuration used here sets $x_{\rm off}=0$.
A scalar produced at $\mathbf r_{0,j}=(x_{0,j},y_{0,j},z_{0,j})$, with $z_{0,j}=z_{{\rm prod},j}$, and traveling along $\widehat{\mathbf n}_j$ follows
\begin{equation}
\mathbf r_j(\ell)=\mathbf r_{0,j}+\ell\widehat{\mathbf n}_j,
\qquad \ell\geq0.
\label{eq:app_scalar_ray}
\end{equation}
Its entry and exit distances are the endpoints of
\begin{equation}
[L_{{\rm in},j},L_{{\rm out},j}]
=
[0,\infty)
\cap
\left[
\frac{z_{\min}-z_{0,j}}{n_{z,j}},
\frac{z_{\max}-z_{0,j}}{n_{z,j}}
\right]
\cap
[\ell_{R,j}^-,\ell_{R,j}^+],
\label{eq:app_ray_cylinder}
\end{equation}
where $\ell_{R,j}^{\pm}$ are the intersections with the cylindrical side wall. The interval is empty if the ray misses the cylinder. For forward trajectories $n_{z,j}>0$, the longitudinal endpoints are ordered as written. A ray parallel to the cylinder axis is accepted transversely only if its fixed transverse position lies within radius $R$.

\subsection{Scalar decay probability and visible acceptance}
\label{app:llp_acceptance}

For scalar laboratory momentum $\mathbf p_j$, the mean decay length is
\begin{equation}
\lambda_j
=\beta_j\gamma_jc\tau_\phi
=\frac{|\mathbf p_j|}{m_\phi}c\tau_\phi.
\label{eq:app_scalar_decay_length}
\end{equation}
The probability to decay between the cylinder entry and exit points is
\begin{equation}
P_{{\rm dec},j}(m_\phi,g)
=
\exp\!\left(-\frac{L_{{\rm in},j}}{\lambda_j}\right)
-
\exp\!\left(-\frac{L_{{\rm out},j}}{\lambda_j}\right).
\label{eq:app_decay_probability}
\end{equation}

For the dimuon search, the two muons are generated isotropically in the scalar rest frame and boosted to the laboratory frame.
For a daughter emitted from $\mathbf r_d$ along $\widehat{\mathbf u}_a$, its intersection with the downstream detector plane is
\begin{equation}
\ell_{a,{\rm det}}
=\frac{z_{\rm det}-z_d}{u_{a,z}},
\qquad
r_{a,{\rm det}}
=
\left|
\mathbf r_{d,\perp}-\mathbf r_{\perp,c}
+\ell_{a,{\rm det}}\widehat{\mathbf u}_{a,\perp}
\right|,
\qquad
\mathbf r_{\perp,c}=(x_{\rm off},0).
\label{eq:app_detector_plane}
\end{equation}
A daughter is geometrically accepted when $u_{a,z}>0$ and $r_{a,{\rm det}}<R_{\rm det}$.
The dimuon factor $\mathcal A_j^{(\mu\mu)}$ requires both tracks to satisfy this condition.
For the diphoton search, the collinear approximation replaces the two resolved photon trajectories by the scalar line of flight: $\mathcal A_j^{(\gamma\gamma)}=1$ when the forward scalar ray from $\mathbf r_d$ intersects the same detector disk and vanishes otherwise.
In the reference calculation the decay point used for acceptance is the midpoint of the in-volume segment, keeping $\mathcal A_j^{(i)}$ independent of $g$.

\subsection{Yield assembly and sensitivity boundary}
\label{app:llp_assembly}

Combining production, propagation, visible branching fraction, and acceptance gives
\begin{align}
N_{\rm sig}^{(i)}(m_\phi,g)
&=
N_{\rm int}\,g^2\mathcal B_i(m_\phi)
\sum_h\sum_{j\in h}
\widehat w_{\phi,j}(m_\phi)
P_{{\rm dec},j}(m_\phi,g)
\mathcal A_j^{(i)}
\label{eq:app_signal_yield}\\
&=
N_{\rm int}\,g^2\mathcal B_i(m_\phi)
\sum_{h,j}
w_{h,j}\frac{\hat{\mathcal B}_h(m_\phi)}{n_h}
\left[
e^{-L_{{\rm in},j}/\lambda_j}
-e^{-L_{{\rm out},j}/\lambda_j}
\right]
\mathcal A_j^{(i)}.
\nonumber
\end{align}
All mass dependence in the production rate and energy spectrum can be tabulated in advance, and a single scalar sample at each mass can be reweighted across the coupling scan.

For each mass and search, the lower and upper sensitivity boundaries are the two crossings of $N_{\rm sig}^{(i)}=3$, when the peak yield exceeds three events. The assumptions and omitted detector effects are specified in \cref{sec:llp_target}.

\section{Performance estimators}
\label{app:metrics}

This appendix collects the statistical definitions and properties of the metric family reported by \toolbench, following the conventions of \cref{sec:toolbench_metrics}.

\label{app:metrics_defs}

An agent attempts an $M$-stage workflow $n$ times. In trial $j$, stage $i$ receives credit $c_{j,i}\in[0,1]$, and its threshold verdict determines whether it gates credit from dependent stages. The general per-trial reach is given in \cref{eq:reach_general}. The forward-LLP rubric combines a shared structural gate with independently graded scientific quantities: the width stage is binary but non-gating, while the remaining physics stages receive continuous non-gating credit. The symbolic rubrics use a shared answer-validity gate followed by independently scored binary checks. The complete choices are given in appendix~\ref{app:rubrics}.

For a binary sequential workflow, the stage verdicts form a matrix $S\in\{0,1\}^{n\times M}$. Under the absorbing convention, $S_{j,i}=0$ for every stage following trial $j$'s first failure, and the general definition reduces to \cref{eq:reach}. With end-only weights ($\mathbf{w}=\mathbf{e}_M$), per-trial reach becomes the binary end-to-end indicator, while mean reach, reach@$k$, and reach$^k$ reduce to pass rate, pass@$k$, and pass$^k$.

Across $k$ repeated attempts, three summaries characterize the result.
\begin{equation}
\overline{\text{reach}} = \mathbb{E}\!\left[R_j\right],
\qquad
\text{reach@}k = \mathbb{E}\!\left[\max_{j\le k} R_j\right],
\qquad
\text{reach}^k = \mathbb{E}\!\left[\min_{j\le k} R_j\right],
\end{equation}
with $\text{reach}^k \le \overline{\text{reach}} \le \text{reach@}k$; for a binary end-only weighting the outer two reduce to pass@$k$ and pass$^k$ with closed forms $1-(1-p)^k$ and $p^k$, where $p$ is the per-trial pass probability.

For $n\geq k$ independent, identically distributed trials, unbiased combinatorial estimators~\cite{chen2021evaluating,liu2024stablereasoning} average the success indicator over all $\binom{n}{k}$ subsets of the observed trials.
For $c$ successful trials they are
\begin{equation}
\widehat{\text{pass@}k} = 1 - \binom{n-c}{k}\!\Big/\!\binom{n}{k}\,,
\qquad
\widehat{\text{pass}^k} = \binom{c}{k}\!\Big/\!\binom{n}{k}\,.
\label{eq:threevector}
\end{equation}
Their graded generalizations follow from order statistics. After sorting the observed reaches $R_{(1)} \le \cdots \le R_{(n)}$, the estimators are
\begin{equation}
\widehat{\text{reach@}k} = \sum_{i=k}^{n} \frac{\binom{i-1}{k-1}}{\binom{n}{k}}\, R_{(i)},
\qquad
\widehat{\text{reach}^k} = \sum_{i=1}^{n-k+1} \frac{\binom{n-i}{k-1}}{\binom{n}{k}}\, R_{(i)},
\end{equation}
where the coefficients are the probabilities that $R_{(i)}$ is the maximum (respectively minimum) of a uniformly random size-$k$ subset.

\section{Campaign specifications and grading}
\label{app:rubrics}

The domain tools used in both campaigns are available in \heptapod~3.0.0. Complete execution configurations, software provenance, and grading records are retained with the campaign artifacts for inspection through the \toolbench dashboard.

\subsection{Forward-LLP campaign}
\label{app:llp_campaign_config}

The complete benchmark specifications, campaign configurations, and trial records are available in the public \pkgname repository.

\subsubsection{Grading}

\Cref{tab:rubric_llp} records the nine stages of the forward-LLP rubric. The seven nonzero weights sum to one. Once the structural gate passes, each scientific stage contributes independently. The complete check parameters are retained in the benchmark's \texttt{benchmark.yaml}.

\begin{table}[H]
\centering
\small
\renewcommand{\arraystretch}{1.12}
\begin{tabularx}{\textwidth}{>{\raggedright\arraybackslash}p{2.5cm}cc>{\raggedright\arraybackslash}X}
\toprule
Stage & Weight & Gate & Evidence and credit \\
\midrule
Answer validity & $0$ & Yes & Well-formed \path{results/answer.json} and populated \path{results/scan.csv}; failure removes all scientific credit. \\
Forward-flux record & $0$ & No & Records whether a forward-flux artifact is retained; does not affect the score. \\
Widths and lifetime & $0.10$ & No & Binary checks of the dimuon expression, dimuon and diphoton widths at $1\%$ and $5\%$ tolerance, and lifetime probes at $5\%$ with $g^{-2}$ scaling. \\
Channel coverage & $0.10$ & No & Fraction of graded search--mass pairs with a nonzero submitted yield. \\
Yield accuracy & $0.40$ & No & $10^{-D_{N_{\rm sig}}}$ comparing yields at 48 specified combinations of search channel, scalar mass, and coupling. Submitted yields are interpolated in $(\log_{10}g,\log_{10}N_{\rm sig})$ onto the reference couplings. \\
Lower boundary & $0.15$ & No & $10^{-D_{g_{\rm lower}}}$ comparing the reconstructed production-limited coupling with its reference value at each required search channel and scalar mass. \\
Upper boundary & $0.15$ & No & $10^{-D_{g_{\rm upper}}}$ comparing the reconstructed lifetime-limited coupling with its reference value at each required search channel and scalar mass. \\
No spurious reach & $0.05$ & No & Penalizes claimed reach outside the reference-sensitive region. \\
Sensitivity figure & $0.05$ & No & Checks drawing content, axis text, and filled regions in the requested figure. \\
\bottomrule
\end{tabularx}
\caption{Forward-LLP rubric. Distances are defined in \cref{eq:llp_distance}. At the reported threshold, a trial passes when $R_r\geq0.9$.}
\label{tab:rubric_llp}
\end{table}

For the lower- and upper-boundary stages in \cref{tab:rubric_llp}, grading requires reconstructing the region where the submitted signal yield exceeds $N_{\rm sig}=3$. At each required search channel and scalar mass, the scan must contain both the lower and upper crossings of this threshold. A missing crossing causes the boundary check to fail. Partial credit is still calculated using the available complete pairs of crossings, with the lower and upper distances evaluated separately. If no complete pair of crossings is found, the boundary credit is zero.

\subsubsection{Statistical contribution to the reach score}
\label{app:llp_statistical_reach}

To interpret departures from unit reach, we estimate the score reduction that finite simulation statistics alone could produce. We compared two independent two-million-event reference builds to estimate this contribution. At a fixed evaluation point, define the signed difference in dex as $\Delta_X=\log_{10}(X_{\rm agent}/X_{\rm ref})$, where the agent and reference simulations contain $n$ and $n_{\rm ref}=2\times10^6$ events, respectively. The standard deviation $\sigma_{\Delta_X}(n)$ describes the spread of these differences over a hypothetical ensemble of independent simulation pairs. When $n=n_{\rm ref}$ and the sampling procedures are equivalent, the two simulations contribute equally to the variance of their difference.

Our reference comparison provides one such pair with $n=n_{\rm ref}$. From this pair, we calculate the mean absolute dex difference across evaluation points, $\langle|\Delta_X|\rangle$. Assuming comparable sampling uncertainties at those points, this average estimates the ensemble mean absolute difference. For approximately zero-mean Gaussian fluctuations, the mean absolute difference is $\sqrt{2/\pi}$ times the standard deviation, giving
\begin{equation}
 \sigma_{\Delta_X}(n_{\rm ref})\simeq\sqrt{\frac{\pi}{2}}\,\langle|\Delta_X|\rangle.
\end{equation}
The mean absolute differences are approximately $0.00266$ dex for peak yields and $0.00283$ dex for boundaries, with the latter combining the lower and upper edges. These give estimated standard deviations of the pairwise differences of $0.00333$ and $0.00354$ dex, respectively. Correlations between evaluation points limit the precision of these estimates.

For equivalent sampling procedures and small fluctuations, sampling variances scale inversely with the number of simulated events. Changing the agent sample from $n_{\rm ref}$ to $n$ therefore multiplies its contribution to the comparison variance by $n_{\rm ref}/n$, while the reference contribution remains fixed. Since each initially contributes half the comparison variance,
\begin{align}
 \sigma_{\Delta_X}^{2}(n)&\simeq\frac{\sigma_{\Delta_X}^{2}(n_{\rm ref})}{2}
 +\frac{\sigma_{\Delta_X}^{2}(n_{\rm ref})}{2}\frac{n_{\rm ref}}{n},
 \notag\\
 \sigma_{\Delta_X}(n)&\simeq\sigma_{\Delta_X}(n_{\rm ref})\sqrt{\frac{1+n_{\rm ref}/n}{2}}.
\end{align}
The scaling factor equals one when the agent and reference samples have the same size. For a zero-mean Gaussian, the median absolute deviation is $0.6745$ times the standard deviation. With zero systematic differences, this gives a representative statistical distance
\begin{equation}
 D_{{\rm stat},X}(n)\simeq0.6745\,\sigma_{\Delta_X}(n_{\rm ref})\sqrt{\frac{1+n_{\rm ref}/n}{2}}.
\end{equation}
This estimates the scale of the median distance when uncertainties are comparable across evaluation points; correlations between points affect its variation between simulations.

To illustrate how this distance enters the reach score, we approximate the full-scan yield distance by the peak-yield estimate and use the pooled boundary estimate for both edges. If all remaining checks receive full credit, the reported weights give
\begin{align}
 R_{\rm stat}(n)
 &\simeq 0.30+0.40\,10^{-D_{{\rm stat},{\rm peak}}(n)}
              +0.30\,10^{-D_{{\rm stat},{\rm boundary}}(n)} \notag\\
 &\simeq 0.30+0.40\,10^{-0.00159\sqrt{1+(2\times10^6)/n}}
              +0.30\,10^{-0.00169\sqrt{1+(2\times10^6)/n}}.
 \label{eq:llp_statistical_reach}
\end{align}
For $n=200{,}000$--$2{,}000{,}000$, the representative distances are approximately $0.002$--$0.006$ dex, corresponding to a reach of $0.991$--$0.996$. This range calibrates the score reduction from finite simulation statistics under the stated assumptions. Different event weights or sampling procedures can change the precision at fixed event count, and peak-yield fluctuations do not determine the uncertainty throughout the yield scan. The retained artifacts do not support a complete uncertainty estimate for every trial, so this calibration alone cannot establish statistical compatibility of the domain-arm results with the reference. The grader records distances divided by the diagnostic scales $a_X=\langle|\Delta_X|\rangle/\sqrt{2}$, while the accuracy credit $10^{-D}$ uses the unscaled distances.

\subsubsection{Sensitivity to stage weights}

To test how the LLP scores depend on the stage weights, we recomputed them from the recorded stage credits, keeping the checks and answer-validity requirement unchanged. The domain arm retains a higher mean score and lower sample standard deviation for all four systems when the seven scored stages receive equal weight. The same holds when only the yield and two boundary stages receive equal weight, although the number of trials above $R_\star=0.9$ changes. Interestingly, restricting the score to the two boundary stages with equal weight reverses the GPT-5.6 comparison, giving a mean of $0.836$ in the core arm and $0.808$ in the domain arm. The aggregate improvement depends on which observables receive weight, motivating the separate yield and boundary comparisons reported in the main text.

\subsection{Symbolic campaign}
\label{app:symbolic_details}

The eight shared tasks and their requested outputs are listed in \cref{tab:symbolic_tasks}. The complete prompts and numerical parameters are retained with the benchmark definitions.

\begin{table}[H]
\centering
\small
\begin{tabular}{lcc}
\toprule
Task & Numerical widths per trial & Formulae per trial \\
\midrule
$F\to f'V$ & 2 & 2 \\
$F\to f'S$ & 2 & 2 \\
$S\to f\bar f$ & 3 & 3 \\
$S\to\chi\chi$ & 2 & 2 \\
$S\to VV$ & 2 & 2 \\
$V\to f\bar f$ & 2 & 2 \\
$V\to SV'$ & 2 & 2 \\
$V\to VV'$ & 2 & 0 \\
\bottomrule
\end{tabular}
\caption{Requested quantities in the eight shared symbolic tasks. Each configuration requests 170 numerical widths and 150 formulae across its 80 trials. The numerical and formula pass fractions exclude trials without a gradeable answer.}
\label{tab:symbolic_tasks}
\end{table}

\subsubsection{Worked task and grading contract}

The scalar-to-fermion task introduced in \cref{sec:symbolic_campaign} requests widths for scalar Yukawa, pseudoscalar, and chiral couplings. For the scalar Yukawa interaction, the reference is
\begin{equation}
\Gamma(S\to f\bar f)
=\frac{y^2m_S}{8\pi}\left(1-\frac{4m_f^2}{m_S^2}\right)^{3/2}
=23.46~\mathrm{GeV}
\end{equation}
at $y=0.87$, $m_S=1240$ GeV, and $m_f=320$ GeV. The pseudoscalar interaction $-iy\phi\bar\psi\gamma_5\psi$ replaces the $\beta^3$ threshold law by $\beta$, while the chiral interaction $-\phi\bar\psi_1(y_LP_L+y_RP_R)\psi_2+\mathrm{h.c.}$ introduces the interference term $-4y_Ly_Rm_1m_2$. The derivation included with the benchmark verifies the reference values to better than $10^{-10}$ against an independent numerical Dirac-algebra calculation.

\begin{table}[H]
\centering
\small
\renewcommand{\arraystretch}{1.12}
\begin{tabularx}{\textwidth}{>{\raggedright\arraybackslash}p{3.0cm}cc>{\raggedright\arraybackslash}X}
\toprule
Requirement & Weight & Gate & Check \\
\midrule
Answer present & $0.04$ total & Yes & Valid JSON with the three numerical widths and three formula fields. \\
Order of magnitude & $0.04$/part & No & Width agrees within half a decade. \\
Numerical accuracy & $0.06$/part & No & Width agrees within $10\%$. \\
Numerical accuracy & $0.08$/part & No & Width agrees within $1\%$. \\
Formula agreement & $0.14$/part & No & Submitted expression agrees with the reference at fixed pseudo-random choices of masses and couplings within the specified ranges. \\
\bottomrule
\end{tabularx}
\caption{Rubric for \texttt{decay\_S\_to\_FFbar}. The four scientific checks are repeated for each of the three coupling structures, giving total weight one. The requested derivation is retained for inspection but does not enter the score.}
\label{tab:rubric_symbolic}
\end{table}

Once the answer-validity gate passes, the numerical tolerances award cumulative credit. For each coupling structure in this worked task, a width within $1\%$ of the reference also passes the $10\%$ and half-decade checks, earning $0.04+0.06+0.08=0.18$ in numerical credit. Formula agreement is graded independently and contributes a further $0.14$ when its check passes.

\subsubsection{Campaign configuration}

The symbolic campaign uses Claude Haiku~4.5 (\texttt{claude-haiku-\allowbreak4-5-20251001}) with Claude Code 2.1.286 for core trials and 2.1.287 for domain trials. The GPT-5.5 core baseline uses Codex 0.152.1 with medium reasoning effort. All use subscription access and ten seeds per task. The domain arm also includes instructions to use the supplied tools, as described in \cref{sec:symbolic_campaign}.

\clearpage
\section{Supplementary campaign evidence}
\label{app:campaign_diagnostics}

The threshold comparison, numerical sensitivity boundaries, and procedural records below support the forward-LLP results in \cref{sec:results_performance}. Additional diagnostic figures and the peak-yield table are retained with the campaign analysis artifacts.

\begin{figure}[H]
\centering
\includegraphics[width=\linewidth]{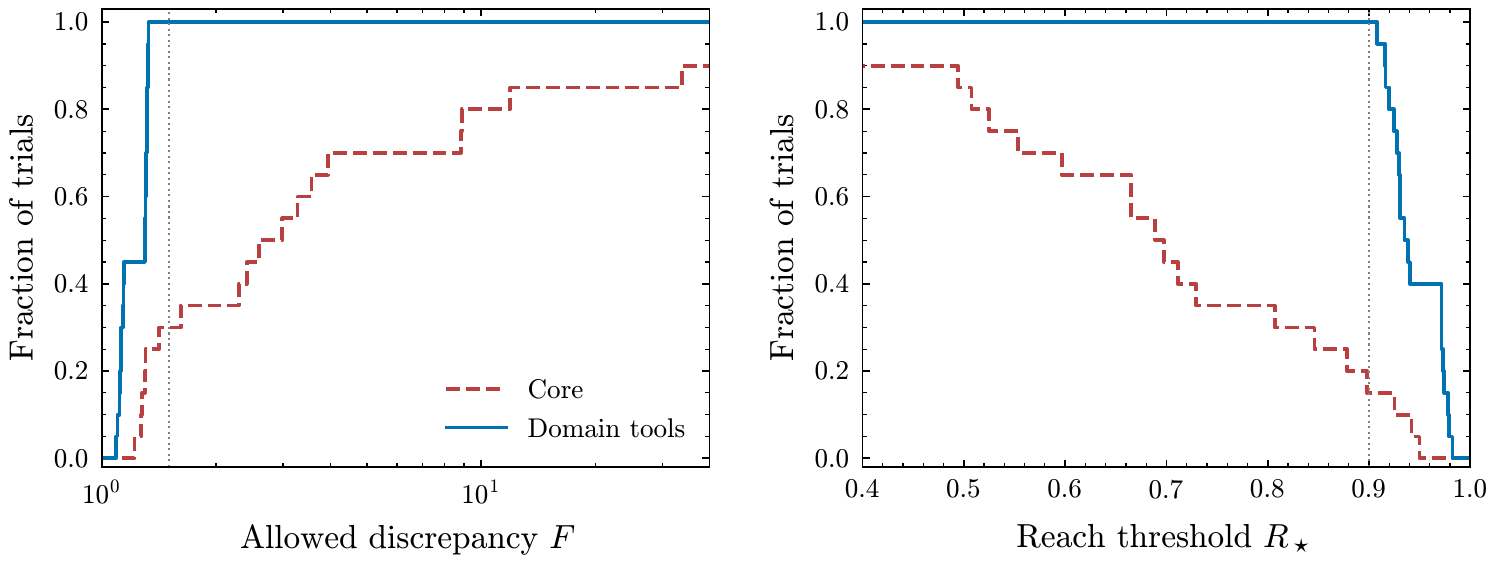}
\caption{Forward-LLP threshold dependence, pooling twenty trials per loadout. Left: Fraction satisfying $F_{\max}\leq F$, with missing distances treated as infinite. Right: Fraction satisfying $R_r\geq R_\star$. Dotted lines mark $F=1.5$ and $R_\star=0.9$. Each step occurs at an observed trial value.}
\label{fig:llp_threshold_sensitivity}
\end{figure}

\begin{figure}[H]
\centering
\includegraphics[width=0.75\linewidth]{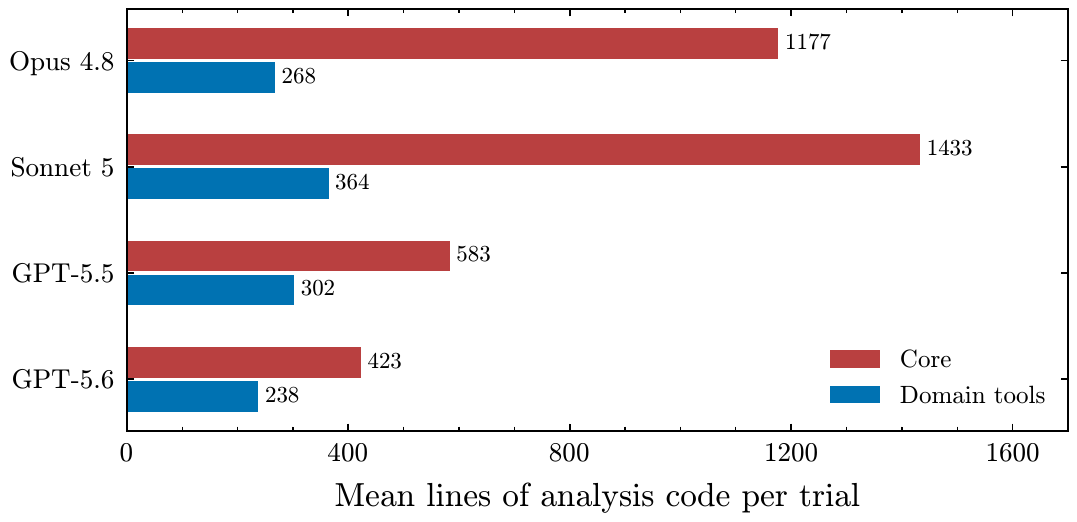}
\caption{Mean lines of agent-authored analysis code per trial, including \textsc{Python} files and code executed inline through the shell. The domain operations themselves are external to this count. Code volume is reported telemetry and does not enter the rubric; a low value can result from successful delegation or incomplete work and must be read together with scientific reach. Here every domain-arm condition also achieves a higher mean reach than its matched core-arm condition.}
\label{fig:llp_code}
\end{figure}

\clearpage

\definecolor{llpA}{RGB}{214,236,214}
\definecolor{llpB}{RGB}{252,244,196}
\definecolor{llpC}{RGB}{252,215,170}
\definecolor{llpD}{RGB}{244,176,170}
\definecolor{llpSpread}{RGB}{160,0,0}
\providecommand{\llpkey}{{\setlength{\fboxsep}{1.5pt}\colorbox{llpA}{$<5\%$}\,\colorbox{llpB}{$5$--$12\%$}\,\colorbox{llpC}{$12$--$41\%$}\,\colorbox{llpD}{$>41\%$}}}

\clearpage
\begin{table}[H]
\centering
\small
\setlength{\tabcolsep}{4pt}
\caption{Boundaries of the sensitivity region in the coupling $g$ at each scalar mass, for the reference and for the domain and core arm of each system. Each entry gives the median over trials, with the minimum and maximum below it. A superscript gives the number of trials that returned a boundary when fewer than five did; a single-trial entry has no range. The lower, production-limited edge is given in units of $10^{-6}$ and the upper, lifetime-limited edge in units of $10^{-3}$. Cell shading gives the deviation of the median from the reference boundary: \protect\llpkey. A range in dark red marks a trial-to-trial spread larger than a factor of two.}
\label{tab:llp_contour}
\begin{tabular}{ll*{5}{c}}
\toprule
& & \multicolumn{5}{c}{$m_\phi$ [GeV]} \\
\cmidrule(lr){3-7}
System & Arm & 0.06 & 0.12 & 0.18 & 0.205 & 0.23 \\
\midrule
\multicolumn{7}{l}{\textit{Lower edge, $g/10^{-6}$}} \\
\midrule
\multicolumn{2}{l}{Reference} & $8.62$ & $5.33$ & $4.35$ & $4.11$ & $0.511$ \\[2pt]
Opus~4.8 & domain & \cellcolor{llpA}$8.62$ & \cellcolor{llpA}$5.33$ & \cellcolor{llpA}$4.38$ & \cellcolor{llpA}$4.10$ & \cellcolor{llpA}$0.511$ \\
 &  & \cellcolor{llpA}{\scriptsize$[8.57,\,10.6]$} & \cellcolor{llpA}{\scriptsize$[5.31,\,7.02]$} & \cellcolor{llpA}{\scriptsize$[4.33,\,5.92]$} & \cellcolor{llpA}{\scriptsize$[4.09,\,5.66]$} & \cellcolor{llpA}{\scriptsize$[0.510,\,0.512]$} \\[2pt]
 & core & \cellcolor{llpC}$10.3^{(3)}$ & \cellcolor{llpC}$6.68$ & \cellcolor{llpC}$5.60$ & \cellcolor{llpC}$5.36$ & \cellcolor{llpA}$0.509$ \\
 &  & \cellcolor{llpC}{\color{llpSpread}{\scriptsize$[5.41,\,11.7]$}} & \cellcolor{llpC}{\color{llpSpread}{\scriptsize$[3.55,\,7.58]$}} & \cellcolor{llpC}{\color{llpSpread}{\scriptsize$[2.98,\,6.47]$}} & \cellcolor{llpC}{\color{llpSpread}{\scriptsize$[2.85,\,6.08]$}} & \cellcolor{llpA}{\color{llpSpread}{\scriptsize$[0.235,\,0.598]$}} \\[2pt]
Sonnet~5 & domain & \cellcolor{llpA}$8.49$ & \cellcolor{llpA}$5.26$ & \cellcolor{llpA}$4.26$ & \cellcolor{llpA}$4.07$ & \cellcolor{llpA}$0.511$ \\
 &  & \cellcolor{llpA}{\scriptsize$[7.55,\,8.59]$} & \cellcolor{llpA}{\scriptsize$[5.03,\,5.35]$} & \cellcolor{llpA}{\scriptsize$[3.74,\,4.35]$} & \cellcolor{llpA}{\scriptsize$[3.55,\,4.09]$} & \cellcolor{llpA}{\scriptsize$[0.485,\,0.511]$} \\[2pt]
 & core & \cellcolor{llpC}$10.1^{(3)}$ & \cellcolor{llpC}$6.55^{(3)}$ & \cellcolor{llpC}$5.45^{(3)}$ & \cellcolor{llpC}$5.22^{(3)}$ & \cellcolor{llpB}$0.472^{(2)}$ \\
 &  & \cellcolor{llpC}{\scriptsize$[9.39,\,17.5]$} & \cellcolor{llpC}{\color{llpSpread}{\scriptsize$[6.15,\,12.8]$}} & \cellcolor{llpC}{\color{llpSpread}{\scriptsize$[5.15,\,11.2]$}} & \cellcolor{llpC}{\color{llpSpread}{\scriptsize$[4.90,\,10.7]$}} & \cellcolor{llpB}{\scriptsize$[0.438,\,0.505]$} \\[2pt]
GPT-5.5 & domain & \cellcolor{llpC}$10.6$ & \cellcolor{llpC}$6.95$ & \cellcolor{llpC}$5.90$ & \cellcolor{llpC}$5.66$ & \cellcolor{llpA}$0.509$ \\
 &  & \cellcolor{llpC}{\scriptsize$[10.6,\,10.7]$} & \cellcolor{llpC}{\scriptsize$[6.92,\,7.07]$} & \cellcolor{llpC}{\scriptsize$[5.85,\,5.90]$} & \cellcolor{llpC}{\scriptsize$[5.62,\,5.67]$} & \cellcolor{llpA}{\scriptsize$[0.507,\,0.510]$} \\[2pt]
 & core & \cellcolor{llpA}$8.51^{(1)}$ & \cellcolor{llpC}$6.35^{(3)}$ & \cellcolor{llpD}$6.31$ & \cellcolor{llpD}$6.82$ & \cellcolor{llpD}$0.359^{(3)}$ \\
 &  & \cellcolor{llpA} & \cellcolor{llpC}{\scriptsize$[5.62,\,11.2]$} & \cellcolor{llpD}{\color{llpSpread}{\scriptsize$[4.42,\,12.4]$}} & \cellcolor{llpD}{\color{llpSpread}{\scriptsize$[4.06,\,10.6]$}} & \cellcolor{llpD}{\color{llpSpread}{\scriptsize$[0.264,\,0.683]$}} \\[2pt]
GPT-5.6 & domain & \cellcolor{llpC}$10.7$ & \cellcolor{llpC}$7.01$ & \cellcolor{llpC}$5.92$ & \cellcolor{llpC}$5.68$ & \cellcolor{llpA}$0.509$ \\
 &  & \cellcolor{llpC}{\scriptsize$[10.6,\,10.7]$} & \cellcolor{llpC}{\scriptsize$[6.95,\,7.07]$} & \cellcolor{llpC}{\scriptsize$[5.90,\,5.93]$} & \cellcolor{llpC}{\scriptsize$[5.65,\,5.69]$} & \cellcolor{llpA}{\scriptsize$[0.506,\,0.513]$} \\[2pt]
 & core & \cellcolor{llpC}$10.3$ & \cellcolor{llpC}$6.50$ & \cellcolor{llpC}$5.67$ & \cellcolor{llpC}$5.33$ & \cellcolor{llpA}$0.514$ \\
 &  & \cellcolor{llpC}{\scriptsize$[9.47,\,15.6]$} & \cellcolor{llpC}{\scriptsize$[6.07,\,10.9]$} & \cellcolor{llpC}{\scriptsize$[5.04,\,9.45]$} & \cellcolor{llpC}{\scriptsize$[4.80,\,9.02]$} & \cellcolor{llpA}{\scriptsize$[0.425,\,0.691]$} \\
\midrule
\multicolumn{7}{l}{\textit{Upper edge, $g/10^{-3}$}} \\
\midrule
\multicolumn{2}{l}{Reference} & $107$ & $21.6$ & $7.29$ & $4.67$ & $0.00619$ \\[2pt]
Opus~4.8 & domain & \cellcolor{llpC}$87.6$ & \cellcolor{llpB}$19.6$ & \cellcolor{llpB}$6.55$ & \cellcolor{llpB}$4.28$ & \cellcolor{llpA}$0.00617$ \\
 &  & \cellcolor{llpC}{\scriptsize$[83.1,\,89.1]$} & \cellcolor{llpB}{\scriptsize$[17.7,\,21.5]$} & \cellcolor{llpB}{\scriptsize$[6.30,\,7.45]$} & \cellcolor{llpB}{\scriptsize$[4.16,\,5.30]$} & \cellcolor{llpA}{\scriptsize$[0.00613,\,0.00634]$} \\[2pt]
 & core & \cellcolor{llpB}$102^{(3)}$ & \cellcolor{llpB}$23.3$ & \cellcolor{llpB}$7.88$ & \cellcolor{llpA}$4.83$ & \cellcolor{llpA}$0.00618$ \\
 &  & \cellcolor{llpB}{\scriptsize$[96.4,\,122]$} & \cellcolor{llpB}{\scriptsize$[21.2,\,28.1]$} & \cellcolor{llpB}{\scriptsize$[7.34,\,10.0]$} & \cellcolor{llpA}{\scriptsize$[4.59,\,6.29]$} & \cellcolor{llpA}{\color{llpSpread}{\scriptsize$[0.00588,\,0.0178]$}} \\[2pt]
Sonnet~5 & domain & \cellcolor{llpC}$94.1$ & \cellcolor{llpC}$18.9$ & \cellcolor{llpB}$6.63$ & \cellcolor{llpB}$4.25$ & \cellcolor{llpA}$0.00619$ \\
 &  & \cellcolor{llpC}{\scriptsize$[88.3,\,126]$} & \cellcolor{llpC}{\scriptsize$[17.8,\,24.3]$} & \cellcolor{llpB}{\scriptsize$[6.22,\,7.52]$} & \cellcolor{llpB}{\scriptsize$[3.98,\,5.47]$} & \cellcolor{llpA}{\scriptsize$[0.00614,\,0.00659]$} \\[2pt]
 & core & \cellcolor{llpB}$99.1^{(3)}$ & \cellcolor{llpA}$21.4^{(3)}$ & \cellcolor{llpA}$7.25^{(3)}$ & \cellcolor{llpA}$4.68^{(3)}$ & \cellcolor{llpB}$0.00664^{(2)}$ \\
 &  & \cellcolor{llpB}{\scriptsize$[76.8,\,104]$} & \cellcolor{llpA}{\scriptsize$[19.1,\,21.8]$} & \cellcolor{llpA}{\scriptsize$[5.83,\,7.94]$} & \cellcolor{llpA}{\scriptsize$[4.01,\,4.96]$} & \cellcolor{llpB}{\scriptsize$[0.00572,\,0.00755]$} \\[2pt]
GPT-5.5 & domain & \cellcolor{llpC}$87.9$ & \cellcolor{llpC}$18.4$ & \cellcolor{llpC}$6.22$ & \cellcolor{llpA}$4.53$ & \cellcolor{llpA}$0.00631$ \\
 &  & \cellcolor{llpC}{\scriptsize$[85.4,\,88.7]$} & \cellcolor{llpC}{\scriptsize$[18.1,\,24.7]$} & \cellcolor{llpC}{\scriptsize$[5.94,\,6.93]$} & \cellcolor{llpA}{\scriptsize$[4.22,\,4.61]$} & \cellcolor{llpA}{\scriptsize$[0.00614,\,0.00663]$} \\[2pt]
 & core & \cellcolor{llpC}$84.1^{(1)}$ & \cellcolor{llpD}$8.36^{(3)}$ & \cellcolor{llpC}$5.53$ & \cellcolor{llpC}$3.55$ & \cellcolor{llpC}$0.00531^{(3)}$ \\
 &  & \cellcolor{llpC} & \cellcolor{llpD}{\color{llpSpread}{\scriptsize$[7.88,\,16.8]$}} & \cellcolor{llpC}{\color{llpSpread}{\scriptsize$[2.42,\,7.15]$}} & \cellcolor{llpC}{\color{llpSpread}{\scriptsize$[1.56,\,5.01]$}} & \cellcolor{llpC}{\scriptsize$[0.00487,\,0.00555]$} \\[2pt]
GPT-5.6 & domain & \cellcolor{llpC}$87.3$ & \cellcolor{llpC}$17.7$ & \cellcolor{llpC}$6.25$ & \cellcolor{llpC}$4.02$ & \cellcolor{llpA}$0.00624$ \\
 &  & \cellcolor{llpC}{\scriptsize$[84.1,\,91.3]$} & \cellcolor{llpC}{\scriptsize$[17.2,\,19.9]$} & \cellcolor{llpC}{\scriptsize$[5.86,\,7.37]$} & \cellcolor{llpC}{\scriptsize$[3.89,\,5.20]$} & \cellcolor{llpA}{\scriptsize$[0.00607,\,0.00646]$} \\[2pt]
 & core & \cellcolor{llpB}$99.4$ & \cellcolor{llpC}$18.5$ & \cellcolor{llpB}$6.81$ & \cellcolor{llpA}$4.74$ & \cellcolor{llpA}$0.00627$ \\
 &  & \cellcolor{llpB}{\scriptsize$[84.7,\,111]$} & \cellcolor{llpC}{\scriptsize$[18.0,\,21.3]$} & \cellcolor{llpB}{\scriptsize$[6.29,\,8.15]$} & \cellcolor{llpA}{\scriptsize$[4.20,\,5.25]$} & \cellcolor{llpA}{\scriptsize$[0.00572,\,0.00716]$} \\
\bottomrule
\end{tabular}
\end{table}

\begin{figure}[p]
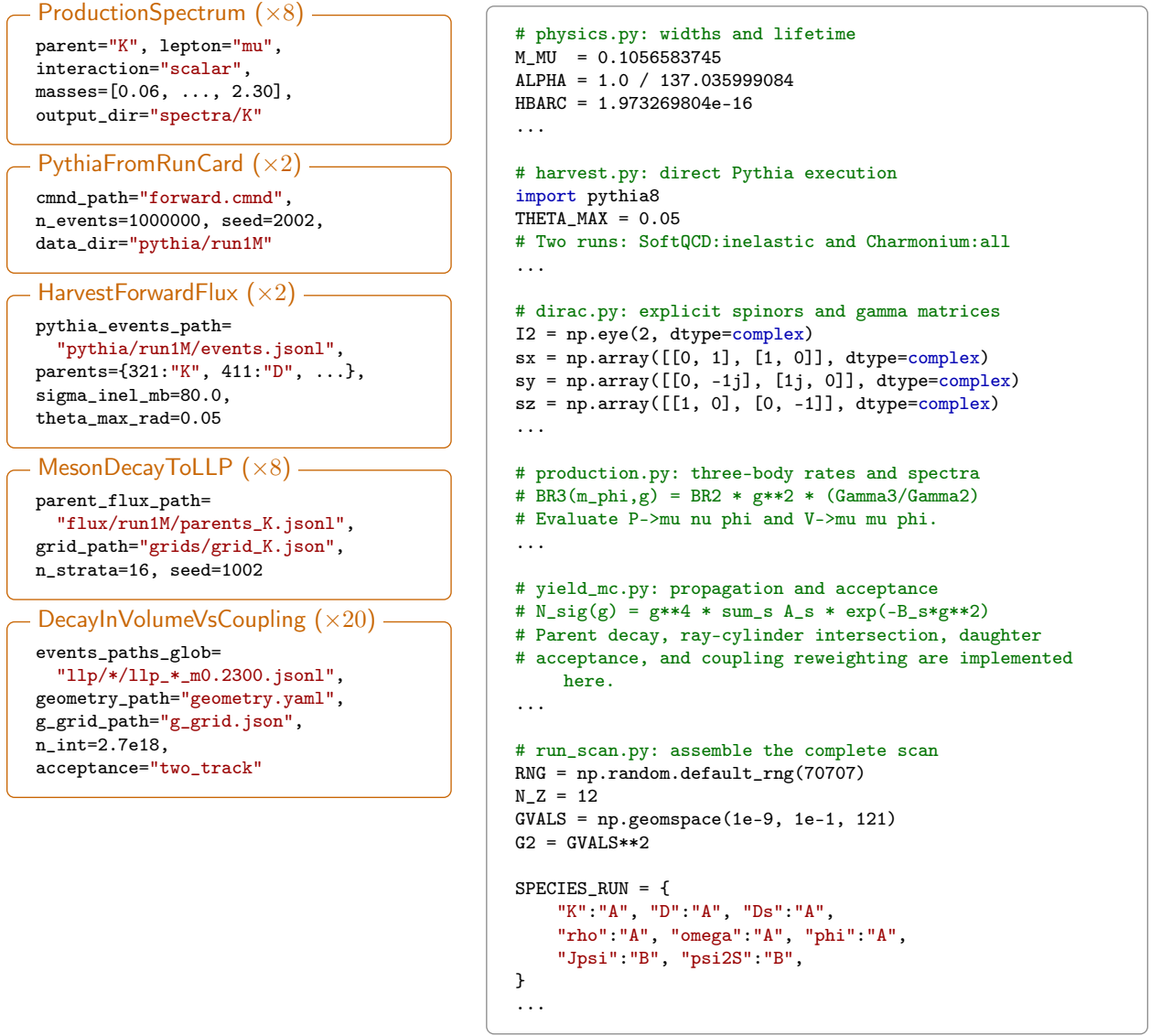

\centering
\begin{minipage}[t]{0.39\linewidth}
\vspace{0pt}
\begin{toolcallpanel}{ProductionSpectrum~($\times 8$)}
\begin{lstlisting}[style=pycode,basicstyle=\ttfamily\fontsize{8.3}{9.5}\selectfont]
parent="K", lepton="mu",
interaction="scalar",
masses=[0.06, ..., 2.30],
output_dir="spectra/K"
\end{lstlisting}
\end{toolcallpanel}

\begin{toolcallpanel}{PythiaFromRunCard~($\times 2$)}
\begin{lstlisting}[style=pycode,basicstyle=\ttfamily\fontsize{8.3}{9.5}\selectfont]
cmnd_path="forward.cmnd",
n_events=1000000, seed=2002,
data_dir="pythia/run1M"
\end{lstlisting}
\end{toolcallpanel}

\begin{toolcallpanel}{HarvestForwardFlux~($\times 2$)}
\begin{lstlisting}[style=pycode,basicstyle=\ttfamily\fontsize{8.3}{9.5}\selectfont]
pythia_events_path=
  "pythia/run1M/events.jsonl",
parents={321:"K", 411:"D", ...},
sigma_inel_mb=80.0,
theta_max_rad=0.05
\end{lstlisting}
\end{toolcallpanel}

\begin{toolcallpanel}{MesonDecayToLLP~($\times 8$)}
\begin{lstlisting}[style=pycode,basicstyle=\ttfamily\fontsize{8.3}{9.5}\selectfont]
parent_flux_path=
  "flux/run1M/parents_K.jsonl",
grid_path="grids/grid_K.json",
n_strata=16, seed=1002
\end{lstlisting}
\end{toolcallpanel}

\begin{toolcallpanel}{DecayInVolumeVsCoupling~($\times 20$)}
\begin{lstlisting}[style=pycode,basicstyle=\ttfamily\fontsize{8.3}{9.5}\selectfont]
events_paths_glob=
  "llp/*/llp_*_m0.2300.jsonl",
geometry_path="geometry.yaml",
g_grid_path="g_grid.json",
n_int=2.7e18,
acceptance="two_track"
\end{lstlisting}
\end{toolcallpanel}
\end{minipage}
\hfill
\begin{minipage}[t]{0.58\linewidth}
\vspace{0pt}
\begin{auditcodepanel}[]
\begin{lstlisting}[style=pycode,basicstyle=\ttfamily\fontsize{8.5}{9.5}\selectfont]
# physics.py: widths and lifetime
M_MU  = 0.1056583745
ALPHA = 1.0 / 137.035999084
HBARC = 1.973269804e-16
...

# harvest.py: direct Pythia execution
import pythia8
THETA_MAX = 0.05
# Two runs: SoftQCD:inelastic and Charmonium:all
...

# dirac.py: explicit spinors and gamma matrices
I2 = np.eye(2, dtype=complex)
sx = np.array([[0, 1], [1, 0]], dtype=complex)
sy = np.array([[0, -1j], [1j, 0]], dtype=complex)
sz = np.array([[1, 0], [0, -1]], dtype=complex)
...

# production.py: three-body rates and spectra
# BR3(m_phi,g) = BR2 * g**2 * (Gamma3/Gamma2)
# Evaluate P->mu nu phi and V->mu mu phi.
...

# yield_mc.py: propagation and acceptance
# N_sig(g) = g**4 * sum_s A_s * exp(-B_s*g**2)
# Parent decay, ray-cylinder intersection, daughter
# acceptance, and coupling reweighting are implemented here.
...

# run_scan.py: assemble the complete scan
RNG = np.random.default_rng(70707)
N_Z = 12
GVALS = np.geomspace(1e-9, 1e-1, 121)
G2 = GVALS**2

SPECIES_RUN = {
    "K":"A", "D":"A", "Ds":"A",
    "rho":"A", "omega":"A", "phi":"A",
    "Jpsi":"B", "psi2S":"B",
}
...
\end{lstlisting}
\end{auditcodepanel}
\end{minipage}
\caption{Two procedural records from paired Opus~4.8 trials at seed 1002. \emph{Left:} Representative arguments for the five domain operations used in the domain-arm trajectory. The multiplicities count all 40 successful domain-operation calls; the remaining 23 recorded calls are core shell, file, or task-management operations. \emph{Right:} Selected excerpts from the 1,020 lines of agent-authored \textsc{Python} and inline code recorded for the core-arm trial. Ellipses mark omitted source. The panels illustrate where the numerical procedure is represented; both arms are graded on the same submitted scientific artifacts.}
\label{fig:llp_audit}
\end{figure}

\newpage
\bibliographystyle{JHEP}
\bibliography{heptabench}

\end{document}